\documentclass[11pt]{article}

\input epsf
\usepackage{latexsym}
\usepackage{amssymb}
\usepackage{amsmath}
\usepackage[dvips]{graphicx}
\usepackage{array}

\def\d3{^{(3)}\nabla}

\begin{document}
\centerline{\Large \bf Effects of helical magnetic fields on the cosmic microwave background}

\vskip 2 cm

\centerline{Kerstin E. Kunze
\footnote{E-mail: kkunze@usal.es} }

\vskip 0.3cm

\centerline{{\sl Departamento de F\'\i sica Fundamental} and {\sl IUFFyM},}
\centerline{{\sl Universidad de Salamanca,}}
\centerline{{\sl Plaza de la Merced s/n, 37008 Salamanca, Spain }}

\vskip 1.5cm

\centerline{\bf Abstract}
\vskip 0.5cm
\noindent
A complete numerical calculation of the temperature anisotropies and polarization of the cosmic microwave background (CMB) in the presence of  a stochastic helical magnetic field is presented which includes the contributions due to scalar, vector and tensor modes. The correlation functions of the magnetic field contributions are calculated numerically including a Gaussian window function to effectively cut off the magnetic field spectrum due to damping.
Apart from parity-even correlations the helical nature of the magnetic field  induces parity-odd correlations between  the E- and B-mode of polarization (EB) as well as between temperature (T) and the polarization B-mode (TB).

\vskip 1cm

\section{Introduction}

The existence of magnetic fields in the universe has long been established by observations on small upto very large scales, that is on the scales of galaxies, galaxy cluster and superclusters \cite{obsgal}. Recent observations even indicate evidence for magnetic fields at truly cosmological scales \cite{obscosmag}.
Over recent years there has been an increasing interest in cosmological magnetic fields, from generation mechanisms to observational imprints on the cosmic microwave background (CMB) (for recent reviews, e.g., \cite{reviews}). 

For a primordial magnetic field to serve as an initial seed field to explain the galactic magnetic field, it is assumed to be generated either in the very early universe, such as due to the amplification of perturbations in the electromagnetic field during inflation (e.g. \cite{tw}), or after inflation has ended, e.g., during a phase transition such as the electroweak or QCD phase transitions in which case it has been shown that the resulting magnetic fields are helical \cite{pt}.  The generation of helical magnetic fields during inflation has been studied in \cite{infhel}. However, in any case it is natural to assume that the magnetic field existed long before initial conditions are set for the evolution of the perturbations determining the temperature anisotropies and polarization of the CMB. Thus observations of the CMB could in principle be used to put limits on the parameters of a primordial magnetic field. The aim here is to calculate the angular power spectra of the temperature anisotropies and polarization for a Gaussian stochastic helical magnetic field.  In this case the magnetic field two-point correlation contains a symmetric and an asymmetric part. Each of which are characterized by one amplitude and one spectral index. The helical part results in non-vanishing correlations of the temperature anisotropies and the polarization B-mode on the one hand and of the polarization E-mode and B-mode of the CMB on the other hand. These have been first considered in \cite{pvw}, where the vector modes were studied. The spectra of the CMB anisotropies due to the tensor mode have been treated in \cite{cdk} and vector and tensor modes in \cite{kr}.  Moreover, the helical part also contributes to all other correlation functions, that is the temperature and polarization E-mode and B-mode autocorrelation functions as well as the temperature polarization E-mode cross correlation of the CMB. Here  a consistent numerical treatment is presented taking into account the magnetic field in the initial conditions and the evolution equations as well as calculating numerically the correlation functions encoding the contribution of the magnetic field.  Using a Gaussian window function the magnetic field spectrum is effectively cut-off at a damping scale.
Moreover, the angular power spectra of the temperature anisotropies and polarization of the CMB are calculated for scalar, vector and tensor modes. 

The calculation of the CMB temperature anisotropies and polarization requires to solve the evolution equations of the perturbation equation of the geometry and matter components together with the Boltzmann equations for photons and neutrinos. Since the first numerical program COSMICS \cite{cosmics} to do this 
and with a significant improvement in speed using the line-of-sight integration in CMBFAST \cite{cmbfast}
there has been a steady evolution of numerical codes such as CAMB \cite{camb}, CMBEASY \cite{cmbeasy}  and CLASS \cite{class}  being the latest addition.
The effect of a primordial nonhelical magnetic field on the CMB anisotropies has been calculated using different approaches: synchronous gauge and thus  a modified version of CMBFAST in \cite{grant,GK}  or the covariant formalism and a modified version of CAMB in \cite{fin,le}. Finally using the gauge-invariant formalism for the scalar perturbations the CMB anisotropies have been calculated in the presence of a stochastic nonhelical magnetic field using CMBEASY in \cite{kek}.

Here the modified version of CMBEASY \cite{kek} has been  expanded to include firstly the numerical calculation of the correlation functions of a helical stochastic Gaussian magnetic field and secondly a new part to solve the Boltzmann equation and calculate the CMB temperature anisotropies and polarization for vector and tensor modes. These are calculated using the total angular momentum approach of Hu and White \cite{hw}. In section \ref{sect2} the decomposition of the magnetic field contribution for scalar, vector and tensor modes is described. Section \ref{sect3} is devoted to the calculation of the relevant correlation functions of  the contribution of a Gaussian stochastic helical magnetic field. As it will be shown there are contributions due to the helical part of the spectrum to the correlation functions for all three modes, namely, scalar, vector and tensor modes. In section \ref{sect4} the perturbation equations for the vector and tensor modes in the presence of a magnetic field are presented in the gauge-invariant formalism \cite{ks}. The corresponding equations for the scalar mode can be found in \cite{kek}. Moreover, the initial conditions are presented. 
 In section \ref{sect5} results of the numerical calculation of the angular power spectra determining the autocorrelation
and cross correlation functions of the temperature (T) and polarization, that is the E-mode (E) and B-mode (B),  of the CMB are presented. Due to the helical nature of the magnetic field apart from the autocorrelation functions of T, E and B and the temperature polarization E-mode cross correlation TE there are also non vanishing correlations between E and B as well as  T and B. The latter one, in particular, is compared for a choice of parameters 
with the WMAP7 data \cite{wmap7}. Section \ref{sect6} contains the conclusions.

\section{Decomposition of the magnetic field contribution}
\label{sect2}
\setcounter{equation}{0}
In a flat Friedmann-Robertson-Walker the lab frame is defined locally by choosing lab coordinates such that $dt=ad\tau$ and $d\vec{r}=ad\vec{x}$ \cite{sb}.
The magnetic field is treated in the lab frame in which it is related to the Maxwell tensor $F_{\mu\nu}$ by
\begin{eqnarray}
B_i(\vec{x},\tau)=\frac{1}{2a^2}\sum_{j,m}\epsilon_{ijm}F_{jm}, 
\end{eqnarray}
where $\epsilon_{ijm}$ is the totally  antisymmetric symbol, with $\epsilon_{123}=1$

The energy-momentum tensor of the electromagnetic field measured by the fundamental observer 
has the form of an imperfect fluid \cite{tsagas}
\begin{eqnarray}
T_{\alpha\beta}=\left(\rho+p\right)u_{\alpha}u_{\beta}+pg_{\alpha\beta}+2u_{(\alpha}q_{\beta)}+\pi_{\alpha\beta}
\end{eqnarray}

where $u^{\alpha}=a^{-1}\delta^{\alpha}_0$ is the four-velocity of the fluid and $u_{\alpha}u^{\alpha}=-1$. The heat flux $q^{\alpha}$ vanishes in a pure magnetic case since it is determined by the Poynting vector. The  magnetic energy density $\rho_{\rm B}$, pressure $p_{\rm B}$ and anisotropic stress 
$\pi_{\rm (B)\;\alpha\beta}$ in the lab frame are given by \cite{tsagas},
\begin{eqnarray}
\rho_{\rm B}=\frac{\vec{B}^2(\vec{x},\tau)}{2},\hspace{1.8cm}p_{\rm B}=\frac{1}{3}\rho_{\rm B},
\hspace{1.8cm}
\pi_{{\rm (B)}\;ij}=-B_i(\vec{x},\tau)B_j(\vec{x},\tau)+\frac{1}{3}\vec{B}^2(\vec{x},\tau)\delta_{ij},
\label{p1}
\end{eqnarray}
where the anisotropic stress has only non-vanishing spatial components and the vector notation denotes a spatial 3-vector. 
Furthermore, the term due to the Lorentz force entering the equation of the baryon velocity evolution and in the tight-coupling limit the photon velocity evolution which is derived from $\nabla^{\alpha}T^{\rm (em)}_{\alpha\beta}=-F_{\beta\alpha}J^{\alpha}$ and expressed in terms of quantities in the lab frame yields to
\begin{eqnarray}
\vec{L}(\vec{x},\tau)=a\left(\vec{J}\times\vec{B}\right)(\vec{x},\tau)
\end{eqnarray}
For vanishing electric field the current $\vec{J}$ is given by
$a\vec{J}=\nabla\times\vec{B}$. Therefore the components of the  Lorentz term  takes the form
\begin{eqnarray}
L_j=-\frac{1}{6}\partial_j\vec{B}^2-\sum_{i}\partial_i\pi_{{\rm (B)}\;ij}
\end{eqnarray}

In the following the magnetic field contributions, namely, the energy density, the anisotropic stress tensor and the Lorentz term are  expanded into scalar, vector and tensor 
harmonic functions. Moreover, it is used that $B_i(\vec{x},\tau)=B_i(\vec{x},\tau_0)\left(\frac{a_0}{a(\tau)}\right)^2$ and $\rho_{\gamma}=\rho_{\gamma 0}\left(\frac{a_0}{a}\right)^4$ where the index 0 refers to the present epoch. 
The energy density of the magnetic field is written in terms of the gauge invariant magnetic energy contrast $\Delta_B$ such that
\begin{eqnarray}
\rho_{\rm B}=\rho_{\gamma}\sum_{\vec{k}}\Delta_{\rm B}(\vec{k})Q^{(0)}(\vec{k},\vec{x}),
\end{eqnarray}
where $Q(\vec{k},\vec{x})$ denote a set of scalar harmonic functions satisfying $(\Delta+k^2)Q^{(0)}=0$
(cf.  e.g. \cite{ks}). Moreover $\Delta_{\rm B}(\vec{k})\equiv\Delta_{\rm B}(\vec{k},\tau_0)$ and similarly the dependence on $\tau_0$ is omitted in the following expressions.
The magnetic anisotropic stress is determined by
\begin{eqnarray}
\pi_{(ij)}(\vec{x},\tau)=p_{\gamma}\sum_{m=0,\pm1, \pm 2}\sum_{\vec{k}}\pi_{\rm B}^{(m)}(\vec{k})Q_{ij}^{(m)}(\vec{k},\vec{x}),
\label{pij}
\end{eqnarray}
where $m=0$ denotes the scalar part and $Q_{ij}^{(0)}=k^{-2}Q_{|ij}+\frac{1}{3}Q^{(0)}$,
the vector part is determined by $m=\pm 1$ and $Q_{ij}^{(\pm 1)}=-\frac{1}{2k}\left(Q_{i|j}^{(\pm 1)}+Q_{j|i}^{(\pm 1)}\right)$ and the tensor modes are given by $m=\pm 2$ \cite{hw}.

Expanding the Lorentz term as
\begin{eqnarray}
L_j(\vec{x},\tau)=\sum_{m=0,\pm 1, \pm 2}\sum_{\vec{k}}L^{(m)}(\vec{k})Q^{(m)}_j(\vec{k}, \vec{x})
\end{eqnarray}
and using that (cf., e.g., \cite{ks}) $Q^{(0)}_{|j}=-kQ_j^{(0)}$, $Q_{ij|i}^{(0)}=\frac{2}{3}kQ_j^{(0)}$
for the scalar harmonics, $Q_{ij|i}^{(\pm 1)}=\frac{k}{2}Q_j^{(\pm 1)}$ for the vector harmonics and $Q^{(\pm 2)}_{ij|i}=0$ for the tensor harmonic functions.
Then the only non-vanishing contributions are given by the scalar and vector part
\begin{eqnarray}
L^{(0)}(\vec{k})&=&\frac{\rho_{\gamma}}{3}k\left(\Delta_{\rm B}-\frac{2}{3}\pi_{\rm B}^{(0)}\right),\\
L^{(\pm 1)}(\vec{k})&=&-\frac{\rho_{\gamma}}{6}k\pi_{\rm B}^{(\pm 1)}(\vec{k}).
\label{vLo}
\end{eqnarray}

The magnetic field is written as
\begin{eqnarray}
B_i(\vec{x},\tau_0)=\sum_{\vec{k}}B_i(\vec{k})Q^{(0)}(\vec{k},\vec{x})
\end{eqnarray}
which implies
\begin{eqnarray}
\Delta_{\rm B}(\vec{k})=\frac{1}{2\rho_{\gamma 0}}\sum_{\vec{q}}B_i(\vec{q})B^i(\vec{k}-\vec{q}).
\end{eqnarray}
A convenient representation of the scalar, vector and tensor harmonic 
functions in flat space is given by \cite{Thorne,Rose,hw}
\begin{eqnarray}
Q^{(0)}(\vec{k},\vec{x})&=&e^{i\vec{k}\cdot\vec{x}}\\
Q^{(\pm 1)}(\vec{k},\vec{x})_i&=&\pm\frac{i}{\sqrt{2}}\left(\hat{e}_1\pm i\hat{e}_2\right)_ie^{i\vec{k}\cdot\vec{x}}\\
Q^{(\pm 2)}_{ij}(\vec{k},\vec{x})&=&-\sqrt{\frac{3}{8}}\left(\hat{e}_1\pm i\hat{e}_2\right)_i\otimes
\left(\hat{e}_1\pm i\hat{e}_2\right)_j e^{i\vec{k}\cdot\vec{x}}.
\label{A1}
\end{eqnarray}
The (spatial) coordinate system defined by the unit vectors $\hat{e}_1$, $\hat{e}_2$ and $\hat{e}_3$ is chosen such that $\hat{e}_3$ lies in the direction of $\vec{k}$, thus  $\hat{e}_3=\hat{k}$.
Moreover, in  the helicity basis \cite{cdk}
\begin{eqnarray}
\hat{e}^{\pm}_{\vec{k}}=-\frac{i}{\sqrt{2}}\left(\hat{e}_1\pm i\hat{e}_2\right)
\end{eqnarray}
so that $\hat{e}_{\vec{k}}^\pm\cdot\hat{e}_{\vec{k}}^{\mp}=-1$ and $\hat{e}^{\pm}_{\vec{k}}\cdot\hat{e}_{\vec{k}}^{\pm}=0$ and  $\hat{e}^{\pm}_{\vec{k}}\cdot\hat{k}=0$.
With this choice the scalar, vector and tensor parts of the anisotropic stress are found to be 
\begin{eqnarray}
\pi_{\rm B}^{(0)}(\vec{k})&=&\frac{3}{2\rho_{\gamma0}}\left[\sum_{\vec{q}}\frac{3}{k^2}B_i(\vec{k}-\vec{q})q^iB_j(\vec{q})\left(k^j-q^j\right)-\sum_{\vec{q}}B_m(\vec{k}-\vec{q})B^m(\vec{q})\right],
\label{p0}\\
\pi_{\rm B}^{(\pm 1)}(\vec{k})&=&\mp i\frac{3}{\rho_{\gamma 0}}\sum_{\vec{q}}\left[\left(\hat{e}_{\vec{k}}^{\mp}\right)^iB_i(\vec{k}-\vec{q})B_j(\vec{q})\hat{k}^j+\left(\hat{e}_{\vec{k}}^{\mp}\right)^jB_j(\vec{q})B_i(\vec{k}-\vec{q})\hat{k}^i\right],\\
\pi_{\rm B}^{(\pm 2)}(\vec{k})&=&-\sqrt{\frac{2}{3}}\frac{3}{\rho_{\gamma 0}}\sum_{\vec{q}}\left(\hat{e}^{\mp}_{\vec{k}}\right)^iB_i(\vec{k}-\vec{q})\left(\hat{e}^{\mp}_{\vec{k}}\right)^jB_j(\vec{q}).
\label{p2}
\end{eqnarray}

\section{Helical magnetic fields}
\label{sect3}
\setcounter{equation}{0}

Magnetic helicity plays an important role in the efficiency of magnetic dynamos.
It provides a measure of the topological structure of the magnetic field, in terms of linkage and twists of its field lines.
It is defined as an integral over the volume $V$ by (cf., e.g. \cite{bis1},\cite{bis2})
\begin{eqnarray}
H_M=\frac{1}{V}\int_V \vec{A}\cdot\vec{B} d^3x
\label{hel}
\end{eqnarray}
where $\vec{A}$ is the gauge potential and $\vec{B}=\vec{\nabla}\times\vec{A}$.
The expression for the magnetic helicity (\ref{hel}) is gauge dependent if 
the normal component of the magnetic field $B_n$ does not vanish on the boundary of the volume.
In this case magnetic helicity is a conserved quantity in the limit of large conductivity. Strictly speaking magnetic helicity is not defined if $B_n\neq 0$. However, more general definitions of helicity have been formulated such as a relative helicity which is manifestly gauge-invariant \cite{fa}. Moreover, there are many physical situations where it is not natural to assume that the magnetic field vanishes on the boundary such as in the case of a stellar magnetic field or the magnetic field inside the horizon. 
In \cite{sbra} a different gauge-invariant definition of the magnetic helicity is proposed.
Similarly to magnetic helicity which describes the complexity of the magnetic field structure there exists the concept of kinetic helicity which determine the structure of the velocity field $\vec{v}$ which is important in turbulence. Kinetic helicity is defined by
\begin{eqnarray}
H_K=\int d^3x \vec{v}\cdot\left(\vec{\nabla}\times\vec{v}\right).
\end{eqnarray}
Therefore, sometimes in analogy to the expression of the kinetic helicity, the quantity
\begin{eqnarray}
H_C\equiv\frac{1}{V}\int d^3x \vec{B}\cdot\left(\vec{\nabla}\times\vec{B}\right)
\end{eqnarray}
 is considered as a measure of the magnetic helicity which is gauge invariant, but is not an ideal invariant as pointed out in \cite{bis1}. This form was used, e.g., in \cite{pvw,cdk,kr}. It describes the (electric) current helicity \cite{mb}.

 Assuming the magnetic field to be a homogeneous Gaussian random field with zero mean, the most general form of the two-point correlation function $C_{ij}$ in real space which is invariant under rotations, but not reflections is given by \cite{my2} (cf. \cite{sub1,sigl})
 \begin{eqnarray}
C_{ij}(\vec{x}_1,\vec{x}_2)= C_{ij}(\vec{r})=\left[C_{L}(r)-C_{N}(r)\right]\frac{r_ir_j}{r^2}+C_{N}(r)\delta_{ij}+C_{A}(r)\epsilon_{ijm}\frac{r_{m}}{r},
 \end{eqnarray}
 where $\vec{r}\equiv \vec{x}_1-\vec{x}_2$ and $r=|\vec{r}|$. Moreover, $C_{L}$ and $C_{N}$ are the longitudinal and lateral correlation functions. The function $C_{A}$ describes the asymmetric part which vanishes in the case of homogeneous, isotropic random fields, 
which by definition are also invariant under reflections. In the case of a magnetic field the asymmetric part is related to its helicity. 
The functions $C_{L}$ and $C_N$ are not independent in the case of the two-point function of a stochastic magnetic field, since 
 $\vec{\nabla}\cdot \vec{B}=0$ \cite{my2,sub1,sigl}. This  reduces by one the number of free functions determining the two-point correlation function in Fourier space.
Therefore, in Fourier space the correlation function for a homogeneous, Gaussian  magnetic field reads \cite{my2} (see also, \cite{pvw,cdk,kr})
\begin{eqnarray}
 \langle B_i^*(\vec{k}) B_j(\vec{k}')\rangle=\delta_{\vec{k}\vec{k}'}P_S(k)\left(\delta_{ij}-\hat{k}_i\hat{k}_j\right)+\delta_{\vec{k}\vec{k}'}P_A(k)i\epsilon_{ijm}\hat{k}_m,
 \label{2point}
\end{eqnarray} 
 where $P_S(k)$ is the power spectrum of the symmetric part which is related to the energy density of the magnetic field and $P_A(k)$ is the power spectrum of the asymmetric part related to the helicity of the magnetic field. Moreover a hat indicates the unit vector, so that $\hat{k}_i\equiv\frac{k_i}{k}$. Following \cite{kek} the powers spectra are chosen to be of the form,
 \begin{eqnarray}
 P_S(k,k_m,k_L)&=&A_B\left(\frac{k}{k_L}\right)^{n_S}W(k,k_m)\\
 P_A(k,k_m,k_L)&=&A_H\left(\frac{k}{k_L}\right)^{n_A}W(k,k_m),
 \end{eqnarray}
where $A_B$ and $A_H$ are the amplitudes and $n_S$ and $n_A$ are the spectral indices of the symmetric and antisymmetric parts, respectively. 
The spectra have to satisfy the so-called realizability condition
$|P_A(k)|\leq P_S(k)$ which is basically a consequence of the Schwartz inequality when applied to the average helicity (cf. e.g. \cite{ab}).
Moreover, $k_L$ is a pivot wave number and $k_m$ is the upper cut-off in the magnetic field spectrum due to diffusion of the magnetic field energy density on small scales. It is assumed that this cut-off is the same for the symmetric and asymmetric parts. Furthermore, $W(k,k_m)$ is a window function.
The damping of the magnetic field is determined by the Alfv\'en velocity and the Silk damping scale which leads to an estimate of the maximal wave number \cite{sb} given by (see also \cite{kko})
\begin{eqnarray}
k_m\simeq 200.694\left(\frac{B}{\rm nG}\right)^{-1} {\rm Mpc}^{-1}
\end{eqnarray}
using the values of the best fit $\Lambda$CDM model of WMAP7 $\Omega_b=0.0227 h^{-2}$ and $h=0.714$ \cite{wmap7}.
The window function is assumed to be Gaussian of the form \cite{kek}
\begin{eqnarray}
W(k,k_m)=\pi^{-\frac{3}{2}}k_m^{-3}e^{-\left(\frac{k}{k_m}\right)^2}
\end{eqnarray}
where the normalization is chosen such that $\int d^3kW(k,k_m)=1$. A different choice of window function, namely a step function, was used in \cite{GK}-\cite{le}.

In the continuum limit $\sum_{\vec{k}}\rightarrow\int\frac{d^3k}{(2\pi)^3}$
the magnetic energy density today smoothed over the magnetic diffusion scale is given by
\begin{eqnarray}
\rho_{B0}=\frac{A_B}{4\pi^{\frac{7}{2}}}\left(\frac{k_m}{k_L}\right)^{n_S}\Gamma\left(\frac{n_S+3}{2}\right)
\end{eqnarray}
which is valid for $n_S>-3$,
where we have used that $\rho_B=\langle\vec{B}(\vec{x},\tau)^2\rangle/2$.
The average helicity measures $H_{M}$ and $H_{C}$ result in 
\begin{eqnarray}
H_M&=&\frac{A_H}{2\pi^{7/2}k_m}\left(\frac{k_m}{k_L}\right)^{n_A}\Gamma\left(\frac{n_A+2}{2}\right)
\end{eqnarray}
which is valid for $n_A>-2$
\begin{eqnarray}
H_C&=&\frac{A_Hk_m}{2\pi^{\frac{7}{2}}}\left(\frac{k_m}{k_L}\right)^{n_A}\Gamma\left(\frac{n_A+4}{2}\right),
\end{eqnarray}
which requires $n_A>-4$.
Therefore the amplitude of the spectral function of the  asymmetric part of the two-point function of the magnetic field can be written as
\begin{eqnarray}
A_H=2\pi^{\frac{7}{2}}{\cal H}_B\left(\frac{k_m}{k_L}\right)^{-n_A},
\end{eqnarray}
where 
\begin{eqnarray}
{\cal H}_B=\left\{
\begin{array}{lr}
H_M k_m/\Gamma\left(\frac{n_A+2}{2}\right)& {\rm magnetic \;\;\; helicity}\\
H_Ck_m^{-1}/\Gamma\left(\frac{n_A+4}{2}\right)& {\rm current \;\;\; helicity}\\
\end{array}
\right.
\end{eqnarray}
In the numerical solutions we consider the maximal allowed contribution of the asymmetric part of the spectrum, that is for $n_A-n_S>0$, the condition $P_A(k_{max})=P_S(k_{max})$ is imposed, where $k_{max}$ is the maximal wave number considered. 
In the opposite case, for $n_A-n_S<0$, the condition $P_A(k_{min})=P_S(k_{min})$ is imposed, where $k_{min}$ is the minimal wave number considered in the numerical solution.
This leads to
\begin{eqnarray}
\left(\frac{{\cal H}_B}{\rho_{\gamma 0}}\right)^2=\left(\frac{\rho_{B0}}{\rho_{\gamma 0}}\right)^2\frac{4}{\Gamma^2\left(\frac{n_S+3}{2}\right)}\left(\frac{q}{k_m}\right)^{2(n_S-n_A)},
\label{hb}
\end{eqnarray}
where $q=k_{max}$ ($k_{min}$)  for $n_A-n_S>0$  ($<0$). Therefore, the larger the 
absolute value of the difference between the spectral indices of the asymmetric and the symmetric part of the magnetic field spectrum the stronger the  helical contribution is suppressed.

The two-point correlation functions of two random fields $F$ and $G$ in $k$-space can be written in terms of the dimensionless spectrum ${\cal P}_{FG}$ as
\begin{eqnarray}
\langle F^*_{\vec{k}}G_{\vec{k}'}\rangle=\frac{2\pi^2}{k^3}{\cal P}_{FG}(k)\delta_{\vec{k},\vec{k}'}.
\label{dsp}
\end{eqnarray}

All correlation functions will be calculated  in the continuum limit.
The autocorrelation function of the magnetic energy density contrast is found to be
\begin{eqnarray}
{\cal P}_{\Delta_{\rm B}\Delta_{\rm B}}(k,k_m)&=&\frac{1}{\left[\Gamma\left(\frac{n_S+3}{2}\right)\right]^2}\left(\frac{\rho_{B,0}}{\rho_{\gamma,0}}\right)^2\left(\frac{k}{k_m}\right)^{2(n_S+3)}e^{-\left(\frac{k}{k_m}\right)^2}
\int_0^{\infty} dz z^{n_S+2}e^{-2\left(\frac{k}{k_m}\right)^2z^2}
\nonumber\\
&&
\int_{-1}^1dx
e^{2\left(\frac{k}{k_m}\right)^2zx}\left(1-2zx+z^2\right)^{\frac{n_S-2}{2}}\left(1+x^2-4zx+2z^2\right)
\nonumber\\
&&
-\frac{{\cal H}_B^2}{2\rho_{\gamma,0}^2}\left(\frac{k}{k_m}\right)^{2(n_A+3)}e^{-\left(\frac{k}{k_m}\right)^2}\int_0^{\infty}dz z^{n_A+2}e^{-2\left(\frac{k}{k_m}\right)^2z^2}
\nonumber\\
&&
\int_{-1}^1dx e^{2\left(\frac{k}{k_m}\right)^2zx}\left(1-2zx+z^2\right)^{\frac{n_A-1}{2}}\left(x-z\right),
\end{eqnarray}
which reduces to the known correlation function for non helical magnetic fields (cf., e.g., \cite{kek}).
Moreover, $x\equiv\frac{\vec{k}\cdot\vec{q}}{kq}$ and $z\equiv\frac{q}{k}$.
Therefore, it is found that the contribution due to the asymmetric part of the correlation function of the magnetic field does not vanish. Similar to  the case of the tensor modes, it is the product of two factor involving the helical part which contributes. Behind this is the observation that the product of two odd parity quantities results in one with even parity \cite{cdk}.
The cross correlation function between the  magnetic energy density contrast and the anisotropic stress in the scalar sector is given by
\begin{eqnarray}
{\cal P}_{\Delta_{\rm B}\pi^{(0)}_{\rm B}}(k,k_m)=\frac{3} {\left[\Gamma\left(\frac{n_S+3}{2}\right)\right]^2}\left(\frac{\rho_{B,0}}{\rho_{\gamma,0}}\right)^2\left(\frac{k}{k_m}\right)^{2(n_S+3)}e^{-\left(\frac{k}{k_m}\right)^2}
\int_0^{\infty} dz z^{n_S+2}e^{-2\left(\frac{k}{k_m}\right)^2z^2}
\nonumber\\
\int_{-1}^1dx
e^{2\left(\frac{k}{k_m}\right)^2zx}\left(1-2zx+z^2\right)^{\frac{n_S-2}{2}}
\left(-1+z^2+zx-(1+3z^2)x^2+3zx^3\right)
\nonumber\\
+\frac{3{\cal H}_B^2}{4\rho_{\gamma,0}^2}\left(\frac{k}{k_m}\right)^{2(n_A+3)}e^{-\left(\frac{k}{k_m}\right)^2}\int_0^{\infty}dz z^{n_A+2}e^{-2\left(\frac{k}{k_m}\right)^2z^2}
\nonumber\\
\int_{-1}^1dx e^{2\left(\frac{k}{k_m}\right)^2zx}\left(1-2zx+z^2\right)^{\frac{n_A-1}{2}}\left(z+2x-3zx^2\right).
\end{eqnarray}
The autocorrelation function of $\pi^{(0)}_{\rm B}$ is determined by
\begin{eqnarray}
{\cal P}_{\pi^{(0)}_{\rm B}\pi^{(0)}_{\rm B}}(k,k_m)=\frac{9} {\left[\Gamma\left(\frac{n_S+3}{2}\right)\right]^2}\left(\frac{\rho_{B,0}}{\rho_{\gamma,0}}\right)^2\left(\frac{k}{k_m}\right)^{2(n_S+3)}e^{-\left(\frac{k}{k_m}\right)^2}
\int_0^{\infty} dz z^{n_S+2}e^{-2\left(\frac{k}{k_m}\right)^2z^2}
\nonumber\\
\int_{-1}^1dx
e^{2\left(\frac{k}{k_m}\right)^2zx}\left(1-2zx+z^2\right)^{\frac{n_S-2}{2}}\left(1+5z^2+2zx+(1-12z^2)x^2-6zx^3+9z^2x^4\right)
\nonumber\\
-\frac{9{\cal H}_B^2}{4\rho_{\gamma,0}^2}\left(\frac{k}{k_m}\right)^{2(n_A+3)}e^{-\left(\frac{k}{k_m}\right)^2}\int_0^{\infty}dz z^{n_A+2}e^{-2\left(\frac{k}{k_m}\right)^2z^2}
\nonumber\\
\int_{-1}^1dx e^{2\left(\frac{k}{k_m}\right)^2zx}\left(1-2zx+z^2\right)^{\frac{n_A-1}{2}}\left(4z+2x-6zx^2\right).
\end{eqnarray}

Since the magnetic field is helical and thus its spectral function has an asymmetric part there are  two different, relevant two-point correlation functions for the anisotropic stress of  vector and tensor modes. The symmetric part which determines the angular power spectra of the CMB due to  all parity even or all parity odd modes is given for the vector modes by
\begin{eqnarray}
\langle\pi_B^{(+1)*}(\vec{k})\pi_B^{(+1)}(\vec{k}')+\pi_B^{(-1)*}(\vec{k})\pi_B^{(-1)}(\vec{k}')\rangle=
\frac{2\pi^2}{k^3}\delta_{\vec{k}\vec{k}'}\left[
\frac{72}{\left[\Gamma\left(\frac{n_S+3}{2}\right)\right]^2}\left(\frac{\rho_{B0}}{\rho_{\gamma 0}}
\right)^2\left(\frac{k}{k_m}\right)^{2(3+n_S)}e^{-\left(\frac{k}{k_m}\right)^2}\right.\nonumber\\
\left.\times 
\int_0^{\infty}dz z^{n_s+2}e^{-2\left(\frac{k}{k_m}\right)^2z^2}
\int_{-1}^1dx e^{2\left(\frac{k}{k_m}\right)^2zx}\left(1-2zx+z^2\right)^{\frac{n_s-2}{2}}(1-x^2)
(1+z^2-3zx+2z^2x^2)\right. \hspace{0.5cm}
\nonumber\\
\left.-\frac{18{\cal H}_B^2}{\rho_{\gamma 0}^2}\left(\frac{k}{k_m}\right)^{2(3+n_A)}e^{-\left(\frac{k}{k_m}\right)^2}
\int_0^{\infty}dz z^{n_A+3}e^{-2\left(\frac{k}{k_m}\right)^2z^2}
\right. \hspace{1cm}
\nonumber\\
\left.
\int_{-1}^1dx
e^{2\left(\frac{k}{k_m}\right)^2zx}(1-2zx+z^2)^{\frac{n_A-1}{2}}(1-x^2)
\right].
\hspace{1cm}
\end{eqnarray}
This   correlation function is used to calculate the angular power spectra of the temperature and polarization autocorrelations, that is $C_{\ell}^{TT}$, $C_{\ell}^{EE}$ and $C_{\ell}^{BB}$, and the temperature polarization E-mode cross correlation $C_{\ell}^{TE}$.
The two-point correlation function for the anisotropic stress of vector modes determining the angular power spectra of the CMB due to a mixture of parity even and parity odd modes is found to be
\begin{eqnarray}
&&\langle\pi_B^{(+1)*}(\vec{k})\pi_B^{(+1)}(\vec{k}')-\pi_B^{(-1)*}(\vec{k})\pi_B^{(-1)}(\vec{k}')\rangle
=\frac{2\pi^2}{k^3}\delta_{\vec{k}\vec{k}'}\frac{36}{\Gamma\left(\frac{n_S+3}{2}\right)}
\left(\frac{\rho_{B0}}{\rho_{\gamma 0}}\right)\left(\frac{{\cal H}_B}{\rho_{\gamma 0}}\right)
\left(\frac{k}{k_m}\right)^{6+n_S+n_A}
\nonumber\\
&&\times e^{-\left(\frac{k}{k_m}\right)^2}
\left[
\int_0^{\infty} dz z^{n_S+2}e^{-2\left(\frac{k}{k_m}\right)^2z^2}\int_{-1}^1dx e^{2\left(
\frac{k}{k_m}\right)^2zx}(1-2zx+z^2)^{\frac{n_A-1}{2}}(1-2zx)(1-x^2)\right.\nonumber\\
&&\left. -\int_0^{\infty}dz z^{n_A+3}e^{-2\left(\frac{k}{k_m}\right)^2z^2}
\int_{-1}^1dx e^{2\left(\frac{k}{k_m}\right)^2zx}
(1-2zx+z^2)^{\frac{n_S-2}{2}}(1-2zx)(1-x^2)
\right]
\label{asvec}
\end{eqnarray}
which is used to calculate the parity-odd correlations of the CMB anisotropies, that is $C_{\ell}^{EB}$ and $C_{\ell}^{TB}$.
It is interesting to note that in the case $n_S=n_A+1$ the correlation function (\ref{asvec}) identically vanishes. 

\begin{figure}[h!]
\centerline{\epsfxsize=3.1in\epsfbox{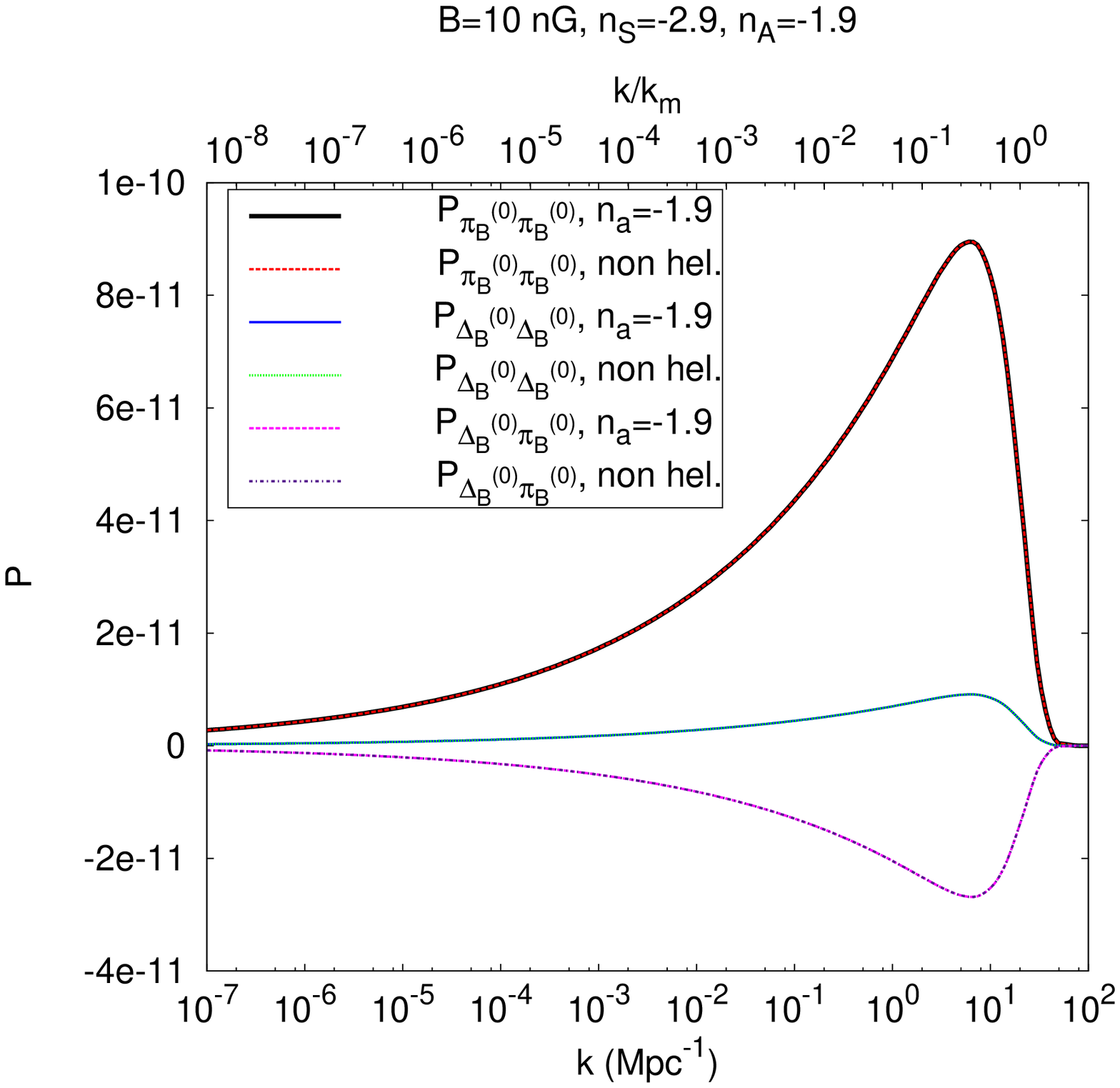}
\hspace{0.1cm}
\epsfxsize=3.1in\epsfbox{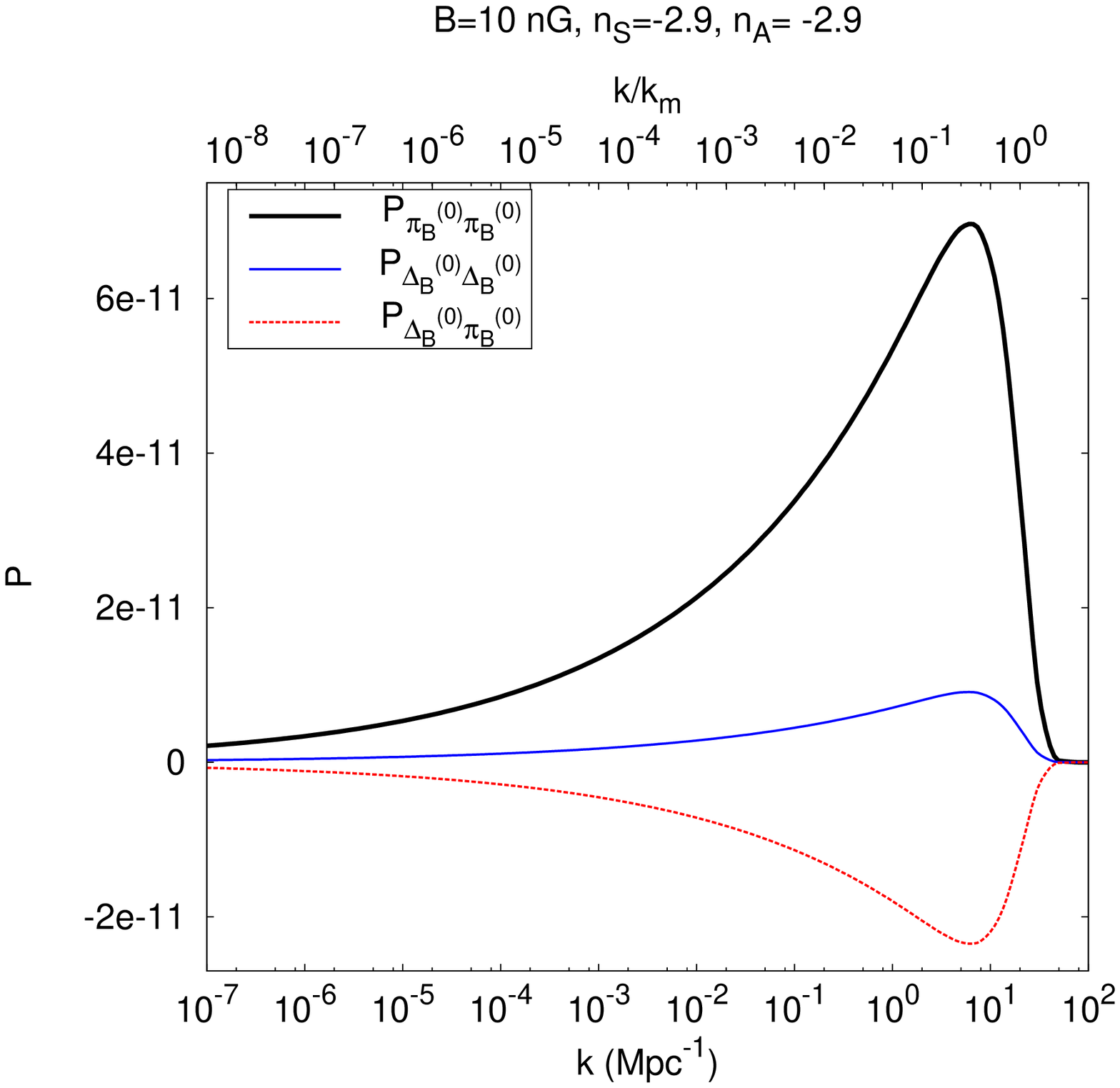}}
\caption{ {\it Left:} The spectra determining the correlation functions for scalar perturbations for $B=10$ nG,  $n_S=-2.9$ and  $n_A=-1.9$  and for comparison the case of a nonhelical magnetic field has been included for  $B=10$ nG,  $n_S=-2.9$. {\it Right: } The spectra determining the correlation functions for scalar perturbations for $B=10$ nG,  $n_S=-2.9$ and  $n_A=-2.9$. The amplitude of the helical component is taken to be the maximal allowed value allowed by the realizability condition. Moreover, the maximal value of the wave number is set to $k_{max}/k_m=100$.}
\label{fig1}
\end{figure}

\begin{figure}[h!]
\centerline{\epsfxsize=3.1in\epsfbox{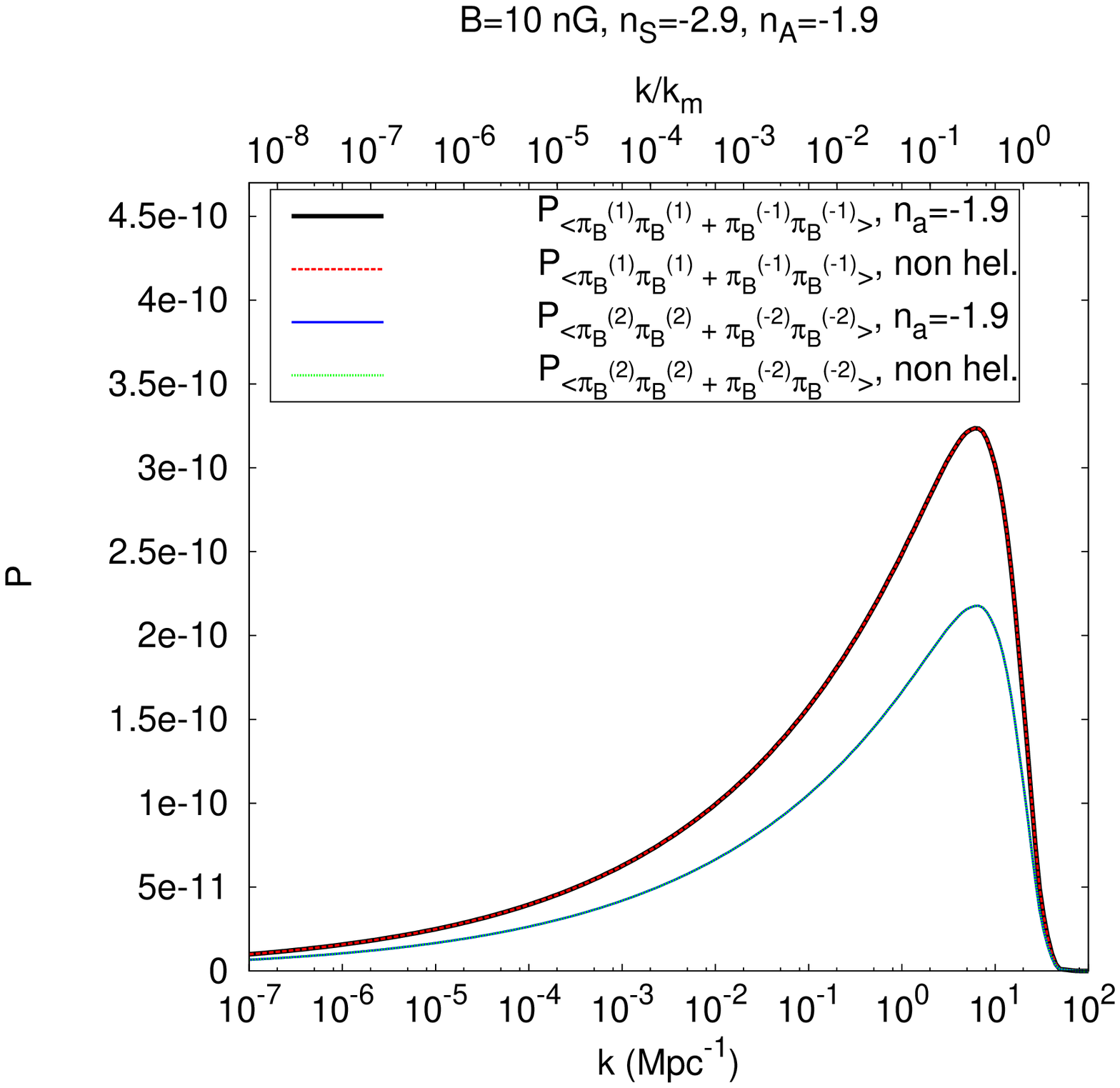}
\hspace{0.1cm}
\epsfxsize=3.1in\epsfbox{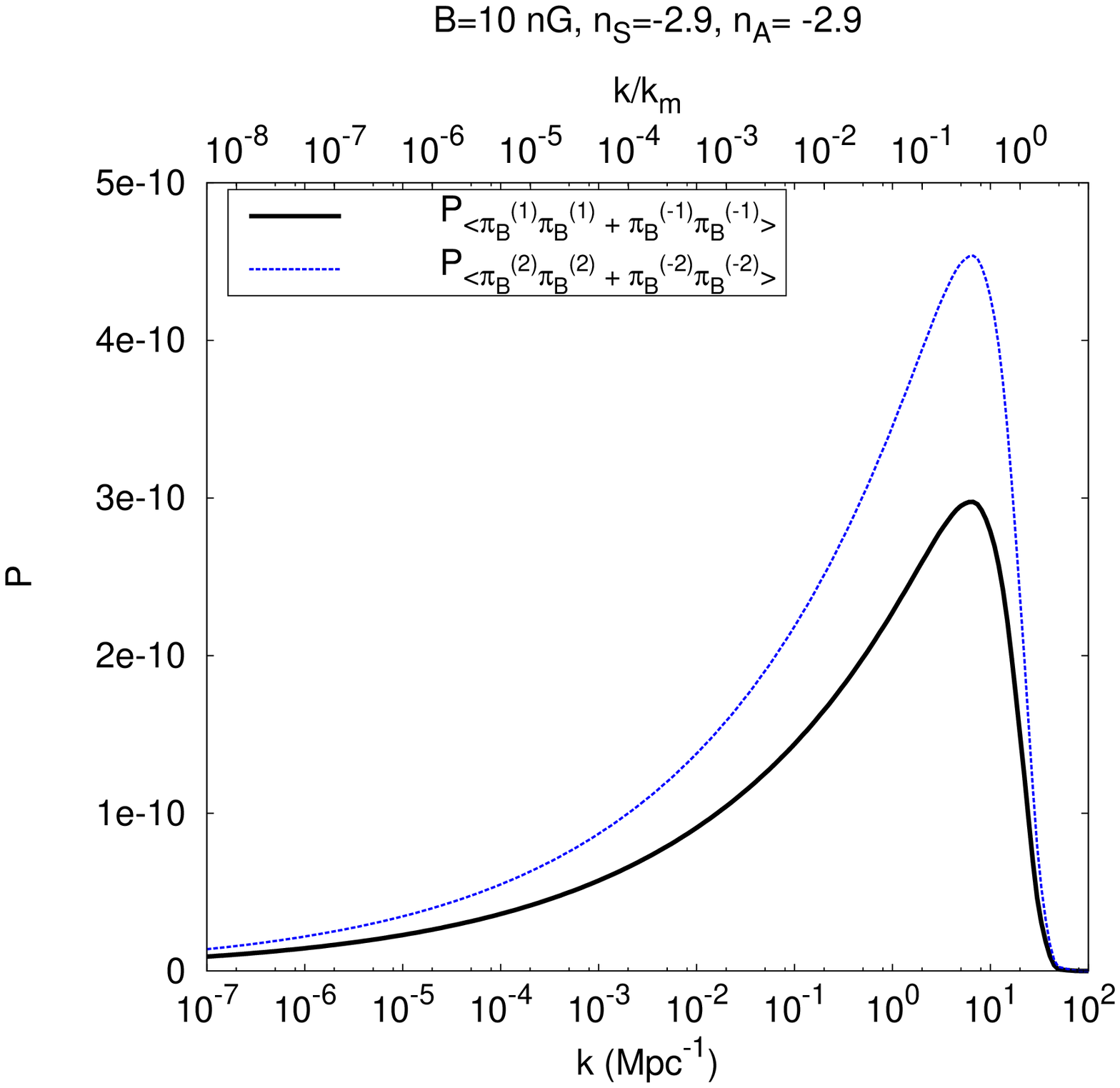}}
\caption{{\it Left:} The spectra determining the even parity correlation functions for vector and tensor modes for $B=10$ nG, $n_s=-2.9$ for $n_A=-1.9$ 
and for comparison the case of a nonhelical magnetic field has been included for  $B=10$ nG,  $n_S=-2.9$.
{\it Right: } The spectra determining the even parity correlation functions for vector and tensor modes  for $B=10$ nG,  $n_S=-2.9$ and  $n_A=-2.9$. The amplitude of the helical component is taken to be the maximal allowed value allowed by the realizability condition. Moreover, the maximal value of the wave number is set to $k_{max}/k_m=100$.}
\label{fig2}
\end{figure}

\begin{figure}[h!]
\centerline{\epsfxsize=3.1in\epsfbox{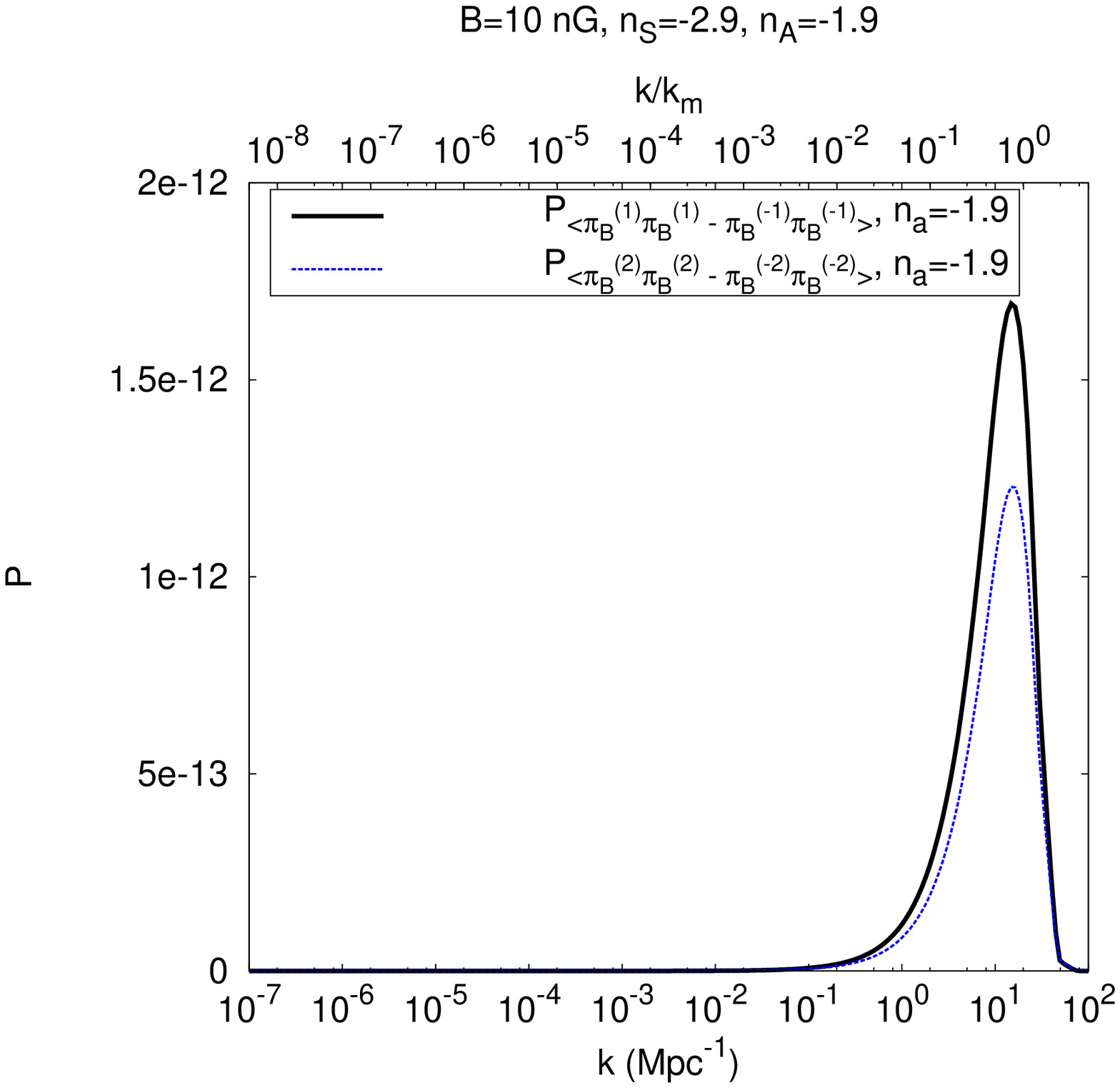}
\hspace{0.1cm}
\epsfxsize=3.1in\epsfbox{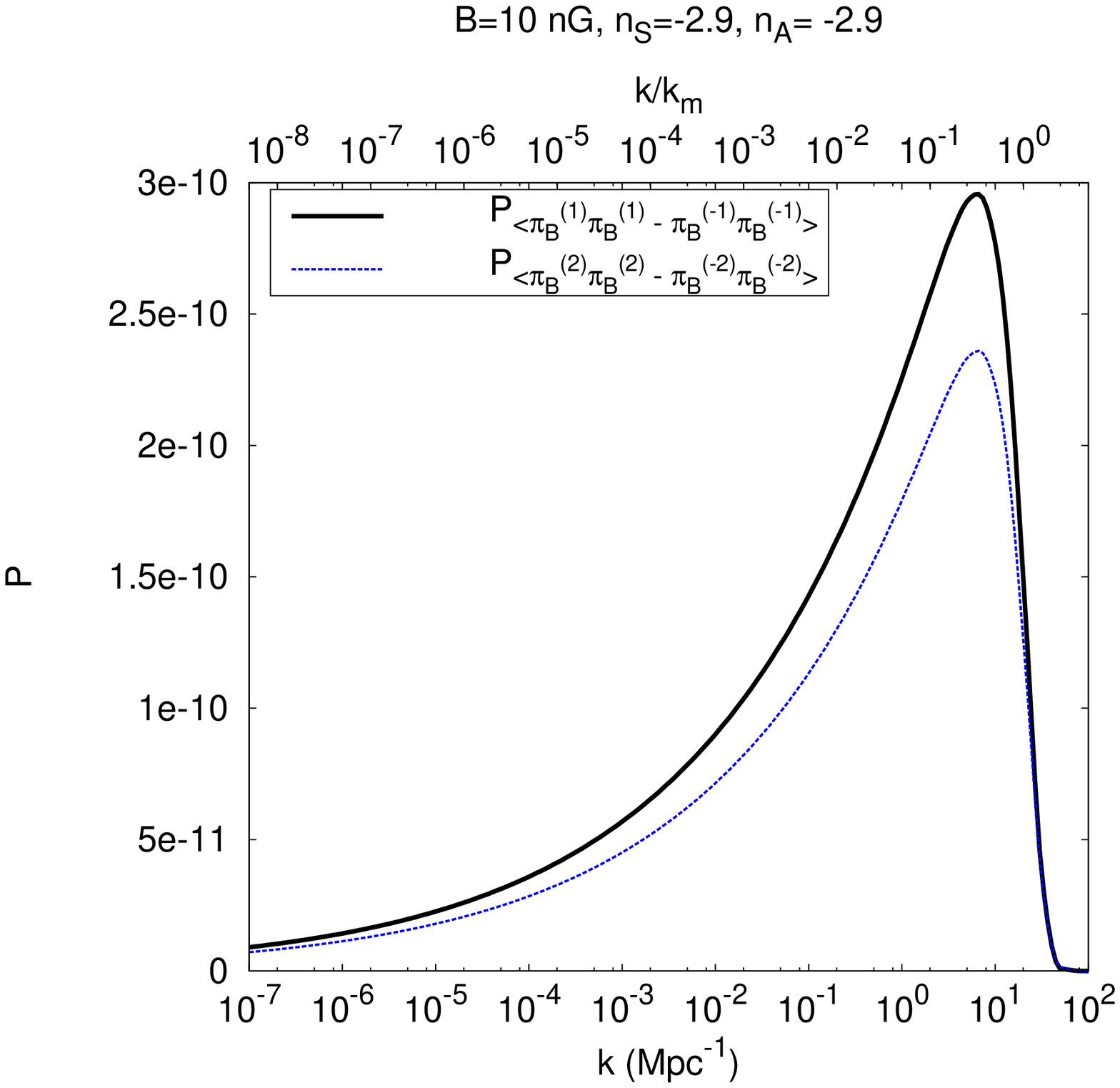}}
\caption{ The spectra determining the odd-parity correlation functions for vector and tensor modes  for $B=10$ nG,  $n_S=-2.9$ and  $n_A=-1.9$  ({\it left})
and $n_A=-2.9$ ({\it right}). The amplitude of the helical component is taken to be the maximal allowed value allowed by the realizability condition. Moreover, the maximal value of the wave number is set to $k_{max}/k_m=100$.}
\label{fig3}
\end{figure}

For the tensor modes the two-point correlation functions of two even-parity modes is given by 
\begin{eqnarray}
&&\langle\pi^{(+2)*}(\vec{k})\pi_B^{(+2)}(\vec{k}')+\pi_B^{(-2)*}(\vec{k})\pi_B^{(-2)}(\vec{k}')\rangle=
\frac{2\pi^2}{k^3}\delta_{\vec{k}\vec{k'}}\left[\frac{24}{\Gamma^2\left(\frac{n_S+3}{2}\right)}
\left(\frac{\rho_{B0}}{\rho_{\gamma 0}}\right)^2\left(\frac{k}{k_m}\right)^{2(3+n_S)}
e^{-\left(\frac{k}{k_m}\right)^2}\right.\nonumber\\
&&\left.\times
\int_0^{\infty}dz z^{n_S+2}e^{-2\left(\frac{k}{k_m}\right)^2z^2}\int_{-1}^1dx
e^{2\left(\frac{k}{k_m}\right)^2zx}(1-2zx+z^2)^{\frac{n_S-2}{2}}(1+x^2)
\left[1-2zx+\frac{z^2}{2}(1+x^2)\right]\right.\nonumber\\
&&\left.
+12\left(\frac{{\cal H}_B}{\rho_{\gamma 0}}\right)^2\left(\frac{k}{k_m}\right)^{2(3+n_A)}
e^{-\left(\frac{k}{k_m}\right)^2}\right.\nonumber\\
&&\left.\times\int_0^{\infty}dz z^{n_A+2}e^{-2\left(\frac{k}{k_m}\right)^2z^2}
\int_{-1}^1dx e^{2\left(\frac{k}{k_m}\right)^2zx}(1-2zx+z^2)^{\frac{n_A-1}{2}}
(1-zx)x
\right]
\end{eqnarray}
which determines $C_{\ell}^{TT}$, $C_{\ell}^{EE}$, $C_{\ell}^{BB}$ and $C_{\ell}^{TE}$ for tensor modes.
The parity-odd correlations of the CMB anisotropies due to tensor perturbations are determined by  
\begin{eqnarray}
\langle\pi_B^{(+2)*}(\vec{k})\pi_B^{(+2)}(\vec{k}')-\pi_B^{(-2)*}(\vec{k})\pi_B^{(-2)}(\vec{k}')\rangle
=\frac{2\pi^2}{k^3}\delta_{\vec{k}\vec{k}'}\frac{12}{\Gamma\left(\frac{n_S+3}{2}\right)}\left(\frac{\rho_{B0}}{\rho_{\gamma 0}}\right)\left(\frac{{\cal H}_B}{\rho_{\gamma 0}}\right)
\left(\frac{k}{k_m}\right)^{6+n_S+n_A}\nonumber\\
\times e^{-\left(\frac{k}{k_m}\right)^2}
\left[2\int_0^{\infty}dz z^{n_A+2}e^{-\left(\frac{k}{k_m}\right)^2z^2}\hspace{2cm}
\right.\nonumber\\
\left.\int_{-1}^1dx
e^{2\left(\frac{k}{k_m}\right)^2zx}x(1-2zx+z^2)^{\frac{n_S-2}{2}}
\left[1-2zx+\frac{z^2}{2}(1+x^2)\right]\hspace{0.4cm}
\right.\nonumber\\
\left. +\int_0^{\infty}dz z^{n_S+2}e^{-2\left(\frac{k}{k_m}\right)^2z^2}\int_{-1}^1
dx e^{2\left(\frac{k}{k_m}\right)^2zx}(1-2zx+z^2)^{\frac{n_A-1}{2}}(1-zx)(1+x^2)
\right]\hspace{0.1cm}
\end{eqnarray}
yielding to non-vanishing  cross correlations between the E- and B- mode $C_{\ell}^{EB}$ and the temperature and B-mode $C_{\ell}^{TB}$. These expressions agree with \cite{cdk}.
The dimensionless spectra (cf.  equation (\ref{dsp})) determining the correlation functions for different choices of $n_S$ and $n_A$ are shown in figure \ref{fig1} for the scalar modes and in figure \ref{fig2} for the even-parity correlations of vector and tensor modes. Finally in figure \ref{fig3} the odd-parity correlations of vector and tensor modes are reported. In the numerical solutions the amplitude of the helical component is taken to be the maximal allowed value allowed by the realizability condition (cf. equation (\ref{hb})).
As can be seen from figure \ref{fig1}, as in the case of nonhelical magnetic fields the magnetic energy density and anisotropic stress are anticorrelated in the case of scalar modes. Moreover, for unequal spectral indices for the symmetric and asymmetric parts of the magnetic field correlation function there is a strong suppression of the helical contribution for the scalar modes. This manifests itself in  the small change between the solutions for $n_A=-1.9$ and a nonhelical magnetic field in figure \ref{fig1} ({\it left}).
As can be appreciated from figure \ref{fig2} for $n_A\neq n_S$ the helical part is strongly suppressed in the even parity correlation functions for vector and tensor modes which is also observed in the case of scalar modes.
The non-vanishing odd-parity correlations for the vector and tensor modes in figure \ref{fig3} constitute a distinctive feature of the helical nature of the magnetic field which lead to the non-vanishing cross correlations $C_{\ell}^{EB}$ and $C_{\ell}^{TB}$ in the CMB.

\section{Perturbation equations}
\label{sect4}
\setcounter{equation}{0}

In this section the perturbation equations and the initial conditions in the presence of a magnetic field for the vector and tensor modes in the gauge-invariant formalism are presented. 
Perturbations are considered around a flat background $ds^2=a^2(\tau)(-d\tau^2+\delta_{ij}dx^idx^j)$, where $a(\tau)$ is the scale factor. 
The corresponding equations for the scalar sector can be found in \cite{kek}. However, we will comment briefly on the initial conditions.
Current numerical codes to calculate the CMB anisotropies, such as \cite{cmbfast,camb,cmbeasy}, set the initial long after neutrino decoupling at which time there is already a non vanishing value of the neutrino anisotropic stress. However, in most models the magnetic field will be generated long before neutrino decoupling. This leads to the magnetic field anisotropic stress being the only contribution to the total anisotropic stress. 
As shown in \cite{kkm,BC,le}  in the case of scalar modes the magnetic anisotropic stress causes the comoving curvature perturbation on superhorizon scales to evolve approaching a constant value shortly after neutrino coupling. Moreover, at this time the neutrino anisotropic stress approaches the compensating solution, in which it balances the magnetic anisotropic stress \cite{mg,GK}. However, the comoving curvature perturbation has acquired an additional contribution due to its
evolution before neutrino decoupling. This is in addition to any primordial curvature perturbation generated, for example, during inflation. The additional contribution is determined by \cite{le}
\begin{eqnarray}
\zeta\simeq-\frac{1}{3}R_{\gamma}\pi_B^{(0)}\ln\frac{\tau_{\nu}}{\tau_B},
\label{pass}
\end{eqnarray}
where $R_{\gamma}\equiv \frac{\Omega_{\nu}}{\Omega_{\gamma}+\Omega_{\nu}}$ and $\tau_B$ is the time of (instantaneous) generation of the magnetic field, which would correspond to the time of the phase transition (e.g. \cite{pt}) or  to reheating if generated during inflation (e.g. \cite{tw}). In  \cite{le} a distinction was made between the compensated magnetic mode with $\zeta=0$ and the adiabatic-like passive mode with $\zeta$ given by equation (\ref{pass}). In \cite{BC} this distinction was not made and here this approach is followed. The initial conditions used in the numerical solution include as a new parameter $\beta\equiv\ln\frac{\tau_{\nu}}{\tau_B}$.

\subsection{Vector perturbations}

There are two gauge-invariant equations \cite{ks}. The amplitude of the shear of the normal vector field to the constant time hypersurface, $\sigma_g^{(1)}$, is determined by
\begin{eqnarray}
\dot{\sigma}^{(1)}_g+2{\cal H}\sigma_g^{(1)}=k\left(\frac{{\cal H}^2}{k^2}\right)\left[
\Omega_{\gamma}\left(\pi_{\gamma}^{(1)}+\pi_B^{(1)}\right)+\Omega_{\nu}\pi_{\nu}^{(1)}\right],
\label{sig}
\end{eqnarray}
where a dot indicates the derivative with respect to conformal time $\tau$ and ${\cal H}=\frac{\dot{a}}{a}$.
The amplitude of the vorticity of the matter velocity field $V^{(1)}$ is gauge-invariant and satisfies
\cite{ks}
\begin{eqnarray}
\dot{V}^{(1)}+(1-3c_s^2){\cal H}V^{(1)}=-\frac{k}{2}\frac{w}{1+w}\pi^{(1)},
\label{vort}
\end{eqnarray}
where $w$ determines the equation of state of the fluid $P=w\rho$ with $P$ the pressure and $\rho$ the energy density and $\pi^{(1)}$ denotes the anisotropic stress. Moreover, $c_s$ is the adiabatic sound speed.
For massless neutrinos ($\nu$) and cold dark matter ($c$) equation (\ref{vort}) yields to
\begin{eqnarray}
\dot{V}^{(1)}_{\nu}&=&-\frac{k}{8}\pi^{(1)}_{\nu}
\label{vnu}\\
\dot{V}^{(1)}_c+{\cal H}V^{(1)}_c&=&0.
\label{vc}
\end{eqnarray}
Deep inside the radiation dominated era photons and electrons are tightly coupled through Thomson scattering and the latter are tightly coupled with the baryons via Coulomb interaction. Thus the baryon, electron and photon fluids are well described by a single fluid. In the tight coupling limit the vorticity fields of photons  ($\gamma$) and baryons ($b$) are determined by
\begin{eqnarray}
\dot{V}_{\gamma}^{(1)}&=&\tau_c^{-1}\left(V_b^{(1)}-V_{\gamma}^{(1)}\right)\\
\dot{V}_b^{(1)}&=&-{\cal H}V_b^{(1)}+R{\tau_c}^{-1}\left(V_{\gamma}^{(1)}-V_b^{(1)}\right)-\frac{R}{8}\pi_B^{(1)},
\label{vortbary}
\end{eqnarray}
where $\tau_c^{-1}$ is the mean free path of photons between scatterings which is determined by the number density of free electrons $n_e$ and the Thomson cross section $\sigma_T$, $\tau_c^{-1}=a n_e\sigma_T$.
Moreover, in the baryon vorticity equation (\ref{vortbary}) $R\equiv\frac{4}{3}\frac{\rho_{\gamma}}{\rho_b}$ and the magnetic field contribution is due to the vector component of the Lorentz force (cf. equation (\ref{vLo})).
At very early times the mean free time between scatterings is much smaller than the Hubble time implying a comparatively large value of $\tau_c^{-1}$. This leads to problems in the numerical integration and was first  solved in the case of scalar perturbations by using an iterative solution at early times and the exact equations at later times \cite{py, maber, doran}.
For the vector perturbations we use a similar approach which results in 
\begin{eqnarray}
\dot{V}_b^{(1)}&=&-\frac{\cal H}{1+R}V_b^{(1)}+\frac{R}{1+R}\left(\dot{\cal V}^{(1)}-\frac{k}{8}\pi_B^{(1)}\right)
\label{vb}\\
\dot{V}_{\gamma}^{(1)}&=&-\frac{R}{1+R}\frac{k}{8}\pi_B^{(1)}-\frac{1}{1+R}\left(\dot{\cal V}^{(1)}+
{\cal H}V_b^{(1)}\right),
\label{vg}
\end{eqnarray}
where the shift $\dot{\cal V}^{(1)}\equiv\dot{V}^{(1)}_b-\dot{V}^{(1)}_{\gamma}$ is determined by 
\begin{eqnarray}
\dot{\cal V}^{(1)}&=&\left[1+2\frac{{\cal H}\tau_c}{1+R}\right]^{-1}\left[\frac{\tau_c}{1+R}
\left(-\frac{\ddot{a}}{a}V_b^{(1)}+\ddot{V}_{\gamma}^{(1)}-\ddot{V}_b^{(1)}-\frac{2{\cal H}}{\tau_c}
\left(V_b^{(1)}-V_{\gamma}^{(1)}\right)\right)\right.
\nonumber\\
&+&\left.
\frac{\dot{\tau}_c}{\tau_c}\left(V_b^{(1)}-V_{\gamma}^{(1)}\right)\right]
\end{eqnarray}
and in the tight-coupling limit the term $\ddot{V}^{(1)}_b-\ddot{V}^{(1)}_{\gamma}$ is neglected.

As pointed out for the scalar modes, initial conditions for the numerical solutions are usually set after neutrino decoupling when the neutrino anisotropic stress is already non vanishing. Therefore before neutrino decoupling the magnetic field is the only source of anisotropic stress \cite{kkm,BC,le}. In the case of the scalar perturbations this leads to an additional contribution to the comoving curvature perturbation as discussed before. 
Thus the only source for the amplitude of the shear of the normal vector field to the constant time hypersurfaces $\sigma_g^{(1)}$ is given by the anisotropic 
stress of the magnetic field, therefore equation (\ref{sig}) implies that at  $\tau_{\nu}$
\begin{eqnarray}
\sigma_g^{(1)}(\tau_{\nu})=\left(\frac{\tau_B}{\tau_{\nu}}\right)^2\sigma_g^{(1)}(\tau_B)+\frac{\Omega_{\gamma}\pi_B^{(1)}}{k\tau_{\nu}}\left(1-\frac{\tau_B}{\tau_{\nu}}\right),
\end{eqnarray}
where $\tau_B$ is the time of generation of the magnetic field. In case it is generated during inflation this time is chosen to coincide with the beginning of the radiation dominated era (see also section \ref{sec4.2}).
After neutrino decoupling, ignoring any photon anisotropic stress, and  neglecting the contribution from the multipole $\ell=4$ the relevant equations from the Boltzmann hierarchy in addition to equation (\ref{vnu}) are given by \cite{hw}
\begin{eqnarray}
\pi_{\nu}^{(1)'}&=&\frac{8}{5}\left(V_{\nu}^{(1)}-\sigma_g^{(1)}\right)-\frac{8\sqrt{24}}{35}N_3^{(1)} \label{n2}\\
N_3^{(1)'}&=&\frac{\pi_{\nu}^{(1)}}{\sqrt{24}} \label{n3}
\end{eqnarray}
where $N^{(1)}_2=\frac{5}{8\sqrt{3}}\pi_{\nu}^{(1)}$.
This gives together with equation (\ref{sig})
\begin{eqnarray}
\pi_{\nu}^{(1)'''}+\frac{3}{x}\pi_{\nu}^{(1)''}+\frac{1}{5}\left(\frac{99}{35}+\frac{8\Omega_{\nu}}{x^2}\right)\pi_{\nu}^{(1)'}+\left(\frac{9}{7}-\frac{8\Omega_{\nu}}{5x^2}\right)\frac{\pi_{\nu}^{(1)}}{x}=\frac{8}{5}\frac{\Omega_{\gamma}\pi_B^{(1)}}{x^3}
\label{pi13}
\end{eqnarray}
where a prime denotes the derivative with respect to $x\equiv k\tau$.
On superhorizon scales $x\ll 1$ equation (\ref{pi13}) is solved by
\begin{eqnarray}
\pi_{\nu}^{(1)}(\tau)=-\frac{\Omega_{\gamma}}{\Omega_{\nu}}\pi_B^{(1)}+\left(k\tau\right)^{-\frac{1}{2}}\left[C_1e^{-\frac{i}{2}\sqrt{32\Omega_{\nu}/5-1}\ln(k\tau)}+C_2e^{\frac{i}{2}\sqrt{32\Omega_{\nu}/5-1}\ln(k\tau)}\right],
\end{eqnarray}
so that for fixed $k$ and late times the compensating solution is approached, $\pi_{\nu}^{(1)}\rightarrow-\frac{\Omega_{\gamma}}{\Omega_{\nu}}\pi_B^{(1)}$.
This implies the vanishing of the righthandside of equation (\ref{sig}) leading to $\sigma_{g}^{(1)}\sim\tau^{-2}$ and therefore suppressing any contribution due to the evolution before neutrino decoupling.
Thus, finally, the compensating magnetic initial conditions for the neutrino anisotropic stress  together with equations 
(\ref{sig}), (\ref{vnu}), (\ref{vc}), (\ref{vb}), (\ref{vg}), (\ref{n2}) and (\ref{n3})  yield the following overall initial conditions  in terms of $x\equiv k\tau$,
\begin{eqnarray}
\sigma_g^{(1)}=\frac{15}{14}\frac{\Omega_{\gamma}\pi_B^{(1)}}{15+4\Omega_{\nu}}x, \hspace{0.5cm}
V_b^{(1)}=V_{\gamma}^{(1)}=-\frac{\pi_B^{(1)}}{8}x, \hspace{0.5cm}
V_{\nu}^{(1)}=\frac{1}{8}\frac{\Omega_{\gamma}}{\Omega_{\nu}}\pi_B^{(1)}x, \hspace{0.5cm}
\pi_{\gamma}^{(1)}=0
\nonumber\\
\pi_{\nu}^{(1)}=\frac{\Omega_{\gamma}}{\Omega_{\nu}}\pi_B^{(1)}\left(-1+\frac{45}{14}\frac{x^2}{15+4\Omega_{\nu}}\right),\hspace{0.5cm}
N_3^{(1)}=\frac{\Omega_{\gamma}}{\Omega_{\nu}}\frac{\pi_B^{(1)}}{\sqrt{24}}\left(-1+\frac{15}{14}\frac{x^2}{15+4\Omega_{\nu}}\right)x.
\end{eqnarray}
which agree with those of \cite{le}.
In deriving these initial conditions it was used that  for the numerical solutions to calculate the CMB anisotropies  these are set deep inside the radiation dominated era when the  baryon-photon fluid is tightly coupled. 

\subsection{Tensor perturbations}
\label{sec4.2}

The metric tensor perturbations are determined by one gauge-invariant amplitude $H_T^{(2)}$ which satisfies \cite{ks}
\begin{eqnarray}
\ddot{H_T}^{(2)}+2{\cal H}\dot{H}_T^{(2)}+k^2H_T^{(2)}={\cal H}^2\left[\Omega_{\gamma}
\left(\pi_{\gamma}^{(2)}+\pi_B^{(2)}\right)+\Omega_{\nu}\pi_{\nu}^{(2)}\right].
\label{eqh}
\end{eqnarray}
It is useful to recall that the initial conditions for the numerical solutions of the Boltzmann equations are set after neutrino decoupling, that is at an initial time $\tau_i>\tau_{\nu}$.
Therefore, as in the case of the scalar and the vector perturbations, the evolution of $H_T^{(T)}$ has to be determined before neutrino decoupling, when the magnetic anisotropic stress is the only source in equation (\ref{eqh}), and after neutrino decoupling when also the anisotropic stress of the neutrinos contributes.
On superhorizon scales, $x=k\tau\ll 1$ using the evolution of $H^{(2)}_T$ during the era before neutrino decoupling, at $\tau_{\nu}$ a regular solution for  $H^{(2)}_T$ is found to be
\begin{eqnarray}
H^{(2)}_T(\tau_{\nu})\simeq H^{(2)}_T(\tau_B)+\Omega_{\gamma}\pi_B^{(2)}\ln\frac{\tau_{\nu}}{\tau_B},
\end{eqnarray}
where $\tau_B$ is the time of generation of the primordial magnetic field if this takes place after the beginning of the radiation dominated era. In this case it is natural to assume $H_T^{(2)}(\tau_B)=0$. If, however, the magnetic field is generated during inflation, then the evolution of $H_T^{(2)}$ during inflation has to be matched to the evolution during the radiation dominated era before neutrino decoupling which is considered here. However, this is beyond the scope of this article and will be considered elsewhere (for work in this direction for the scalar modes see \cite{bcd}). For simplicity, it is assumed that $\tau_B=\tau_{RH}$ the time of reheating when the standard radiation dominated era begins and in the numerical solution $H_T^{(2)}(\tau_B)=0$.
After neutrino decoupling the amplitude of the tensor mode $H_T^{(2)}$ and the neutrino anisotropic stress $\pi_{\nu}^{(2)}$ satisfy \cite{hw},
\begin{eqnarray}
H_T^{(2)''}+\frac{2}{x}H_T^{(2)'}+H_T^{(2)}&=&x^{-2}\left[\Omega_{\gamma}\pi_B^{(2)}+\Omega_{\nu}\pi_{\nu}^{(2)}\right], \label{s1}\\
\pi_{\nu}^{(2)'}&=&-\frac{8}{5}H_T^{(2)'}-\frac{8}{7\sqrt{5}}N_3^{(2)} \label{s2}\\
N_3^{(2)'}&=&\frac{\sqrt{5}}{8}\pi_{\nu}^{(2)}, \label{s3}
\end{eqnarray}
where a prime indicates the derivative with respect to $x=k\tau$ and the neutrino multipole for $\ell=4$ has been neglected.
Equations (\ref{s1}) to (\ref{s3}) can be combined to give a fourth order differential equation
 for $H_T^{(2)}$,
 \begin{eqnarray}
 H_T^{(2)''''}+\frac{6}{x}H_T^{(2)'''}+\left[\frac{8}{7}+\frac{2}{5}\frac{15+4\Omega_{\nu}}{x^2}\right]H_T^{(2)''}+\frac{30}{7x}H_T^{(2)'}+\left(\frac{1}{7}+\frac{2}{x^2}\right)H_T^{(2)}=\frac{1}{7}\frac{\Omega_{\gamma}\pi^{(2)}_B}{x^2}.
 \end{eqnarray}
 On superhorizon scales $x\ll 1$ a regular solution at some time $\tau_i>\tau_{\nu}$,
which is going to be taken the initial time to start the numerical evolution of the Boltzmann code, is given by
\begin{eqnarray}
H_T^{(2)}(\tau_i)&=& H_T^{(2)}(\tau_B)\left[1-\frac{5x^2}{2(15+4\Omega_{\nu})}\right]+
\Omega_{\gamma}\pi^{(2)}_B\ln\frac{\tau_{\nu}}{\tau_B}
\left[1-\frac{5x^2}{2(15+4\Omega_{\nu})}\right]
\nonumber\\
&&+\Omega_{\gamma}\pi_B^{(2)}\frac{5x^2}{28(15+4\Omega_{\nu})}+{\cal O}(x^3),
\end{eqnarray}
 where the solution has been determined by matching the solution before and after neutrino decoupling at $\tau=\tau_{\nu}$.
Similarly, using equations (\ref{s1}) to (\ref{s3}) the evolution of the anisotropic stress of the neutrinos is determined by,
\begin{eqnarray}
\left(1-\frac{2}{x^2}\right)\pi_{\nu}^{(2)''''}&+&\frac{2}{x}\left(1-\frac{4}{x^2}\right)\pi_{\nu}^{(2)'''}
+\left[\frac{8}{7}+\frac{1}{x^2}\left(-\frac{44}{7}+\frac{8\Omega_{\nu}}{5}\right)-\frac{16}{5}\frac{\Omega_{\nu}}{x^4}\right]\pi_{\nu}^{(2)''}\nonumber\\
&+&\frac{2}{35}\frac{1}{x}\left[5-\frac{4\left(5+28\Omega_{\nu}\right)}{x^2}+\frac{112\Omega_{\nu}}{x^4}\right]\pi_{\nu}^{(2)'}\nonumber\\
&+&\left[\frac{1}{7}-\frac{6}{7x^2}+\frac{48\Omega_{\nu}}{5x^4}-\frac{32\Omega_{\nu}}{5x^6}\right]\pi_{\nu}^{(2)}=\frac{16}{5}\left(\frac{2}{x^6}-\frac{3}{x^4}\right)\Omega_{\gamma}\pi_B^{(2)}.
\end{eqnarray}
 On large scales $x\ll 1$ a regular solution is found to be
 \begin{eqnarray}
 \pi_{\nu}^{(2)}(\tau_i)= -\frac{\Omega_{\gamma}}{\Omega_{\nu}}\pi_B^{(2)}+\left[\frac{4}{15+4\Omega_{\nu}}H_T^{(2)}(\tau_B)+\frac{4\Omega_{\gamma}\pi^{(2)}_B}{15+4\Omega_{\nu}}\ln\frac{\tau_{\nu}}{\tau_B}+\frac{15}{14}\frac{\Omega_{\gamma}}{\Omega_{\nu}}\frac{\pi_B^{(2)}}{15+4\Omega_{\nu}}\right]x^2+{\cal O}(x^3)
 \end{eqnarray}
 where equations (\ref{s2}) and (\ref{s3}) have been used to determine the free constant.
 Therefore, the solution approaches the one corresponding to the so-called compensating mode.
 However, since there  are no independent constants multiplying the passive and the compensating mode 
 the two parts of the initial conditions are not treated separately  as in \cite{le}. 
This follows \cite{BC} where  in the case of scalar perturbations  no separation into a passive and a compensating mode was made.  
We close this section by noting that in the numerical solution  the standard tight-coupling approximation is used \cite{le}, in  the notation of \cite{hw}, $\Theta_2^{(2)}=\frac{5}{8}=\pi_{\gamma}^{(2)}=-\frac{4}{3}\tau_C\dot{H}_T^{(2)}$ for the anisotropic stress of the photons, and  
$E^{(2)}_2=-\frac{\sqrt{6}}{4}\Theta_2^{(2)}$, $B_2^{(2)}=0$ for the polarization.

\section{Results}
\label{sect5}

In \cite{kek} the CMB temperature anisotropies and polarization due to scalar perturbations in the gauge-invariant formalism in the presence of a stochastic magnetic field have been calculated using a modified version of CMBEASY \cite{doran}. As opposed to the case considered here, it was assumed that the magnetic field is non helical. 
The calculation of the angular power spectra of the CMB temperature anisotropies and polarization in the presence of a helical stochastic  magnetic field
as described by the two-point correlation function (\ref{2point}) has been done by expanding the numerical code of \cite{kek}. In the case of the scalar perturbations the initial conditions have been changed in order to include the contribution of the magnetic field to the curvature perturbation due to its evolution before neutrino decoupling. Moreover, a new part has been added to the modified version of CMBEASY \cite{kek} to include the numerical solution of the Boltzmann hierarchy and the calculation of the CMB anisotropies and polarization for the vector and tensor modes using the total angular momentum approach of Hu and White \cite{hw}. In the numerical solution here it is not assumed that there is an explicit separation into a magnetic mode and a passive mode as in \cite{le}. The initial conditions for the numerical solution are set long after neutrino decoupling. For scalar and tensor modes  a new parameter, $\beta\equiv\ln\frac{\tau_{\nu}}{\tau_B}$, is included encoding the evolution of the comoving curvature perturbation in the case of the scalar mode and the amplitude of the tensor mode, respectively, after the creation of the magnetic field at $\tau_B$ and before neutrino decoupling at $\tau_{\nu}$.
The vector mode is not affected significantly by the presence of a magnetic field before neutrino decoupling.
The spectrum of the stochastic magnetic field is effectively cut off at the magnetic diffusion scale using a Gaussian window function \cite{kek}.
The contribution of the magnetic field is determined by the two-point correlation functions involving the magnetic energy density and anisotropic stress leading to convolution integrals in Fourier space. Rather than using an approximation, as in previous work, which however, also differs in the sharp cut-off of the magnetic field spectrum as opposed to the Gaussian window function used here, e.g. \cite{GK,fin,le},  these integrals are calculated numerically in the code as was done in \cite{kek} for the non helical case for the scalar mode.
The  angular power spectra describing the CMB temperature anisotropies and polarization are obtained following a treatment similar to the one of correlated isocurvature modes (e.g. \cite{iso}) \cite{le,kek} with the relevant correlation functions being the ones of the magnetic field contribution.

\begin{figure}[h!]
\centerline{\epsfxsize=3.1in\epsfbox{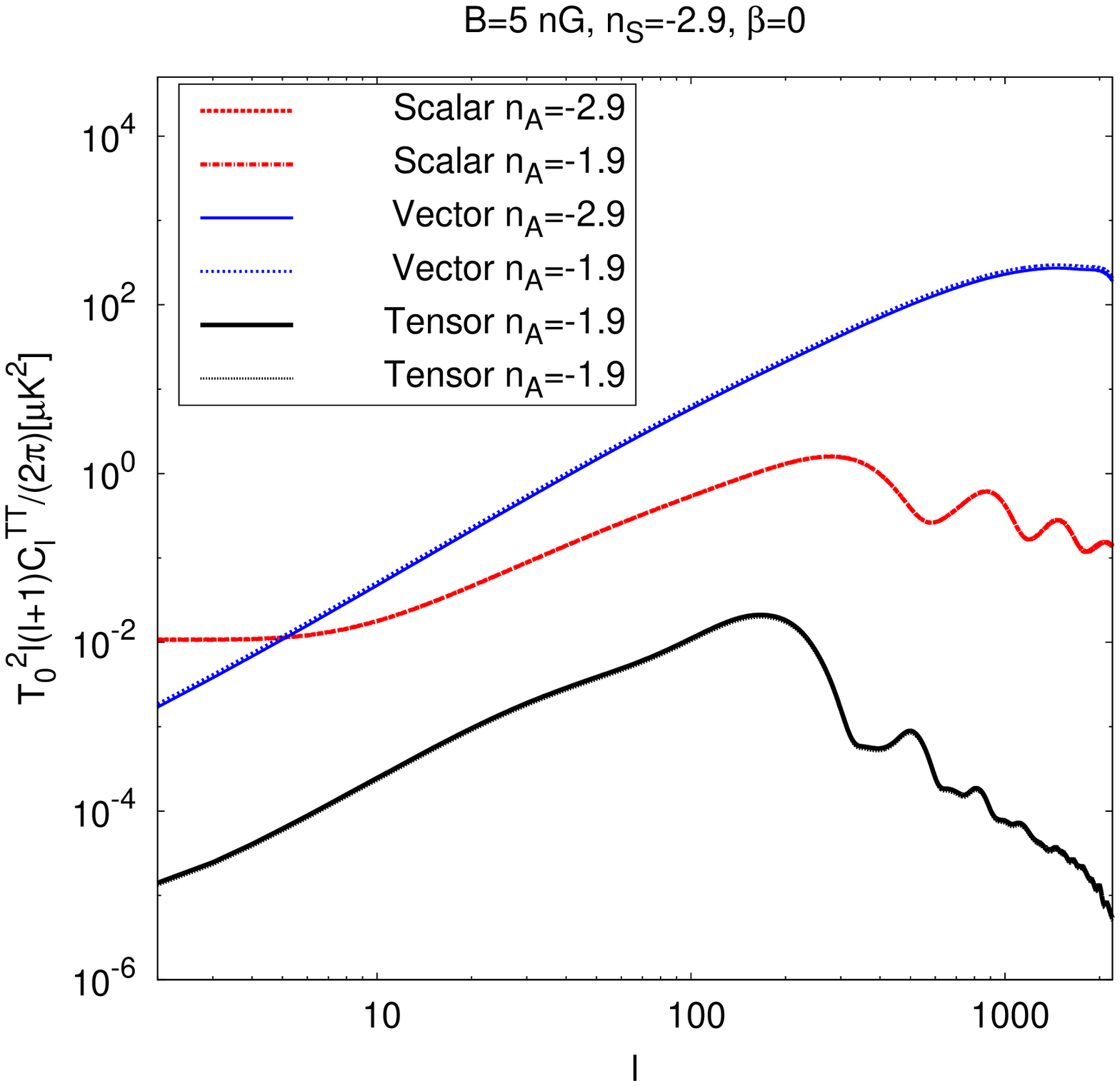}
\hspace{0.2cm}
\epsfxsize=3.1in\epsfbox{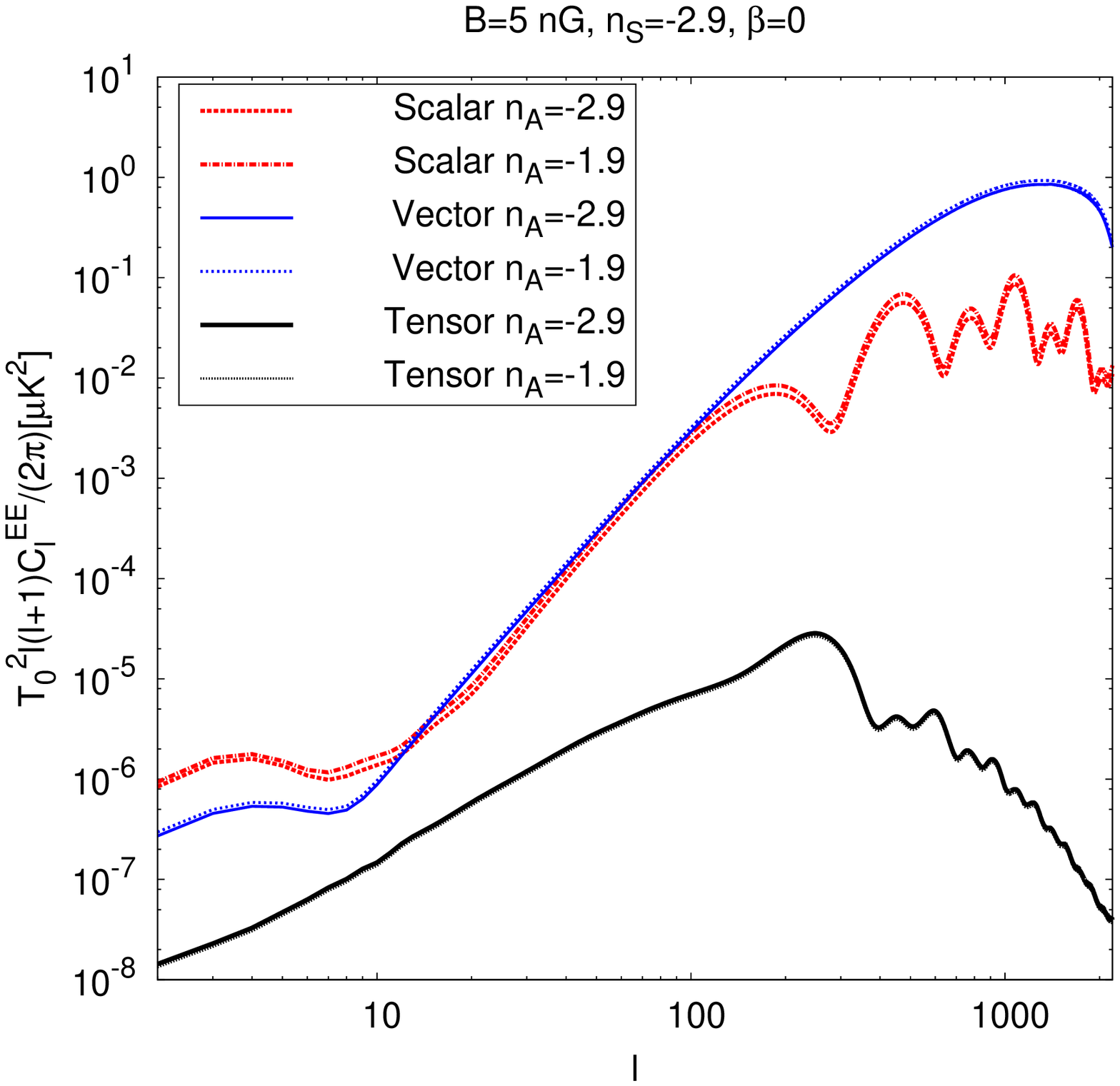}}
\centerline{
\epsfxsize=3in\epsfbox{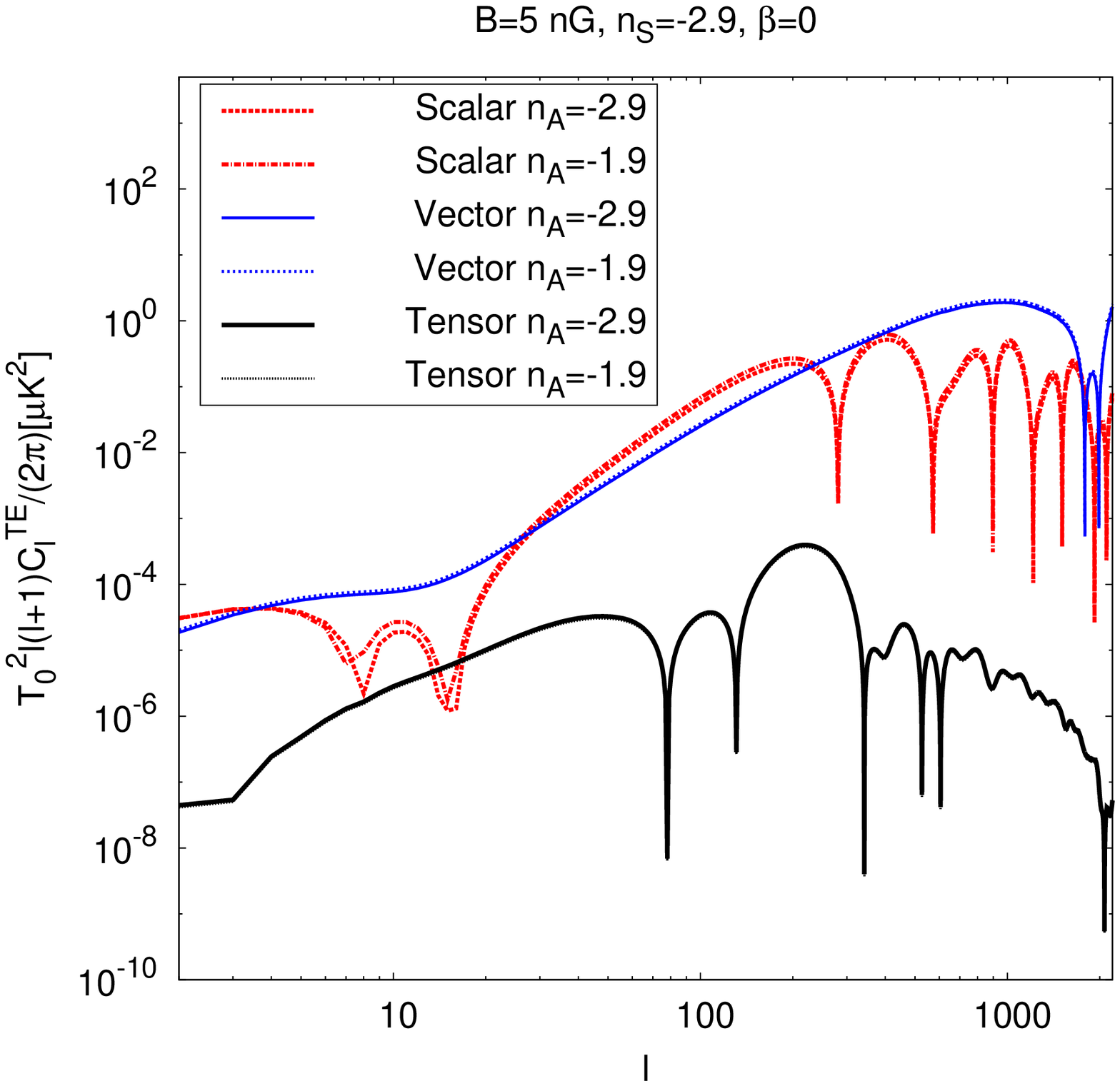}
\hspace{0.2cm}
\epsfxsize=3.1in\epsfbox{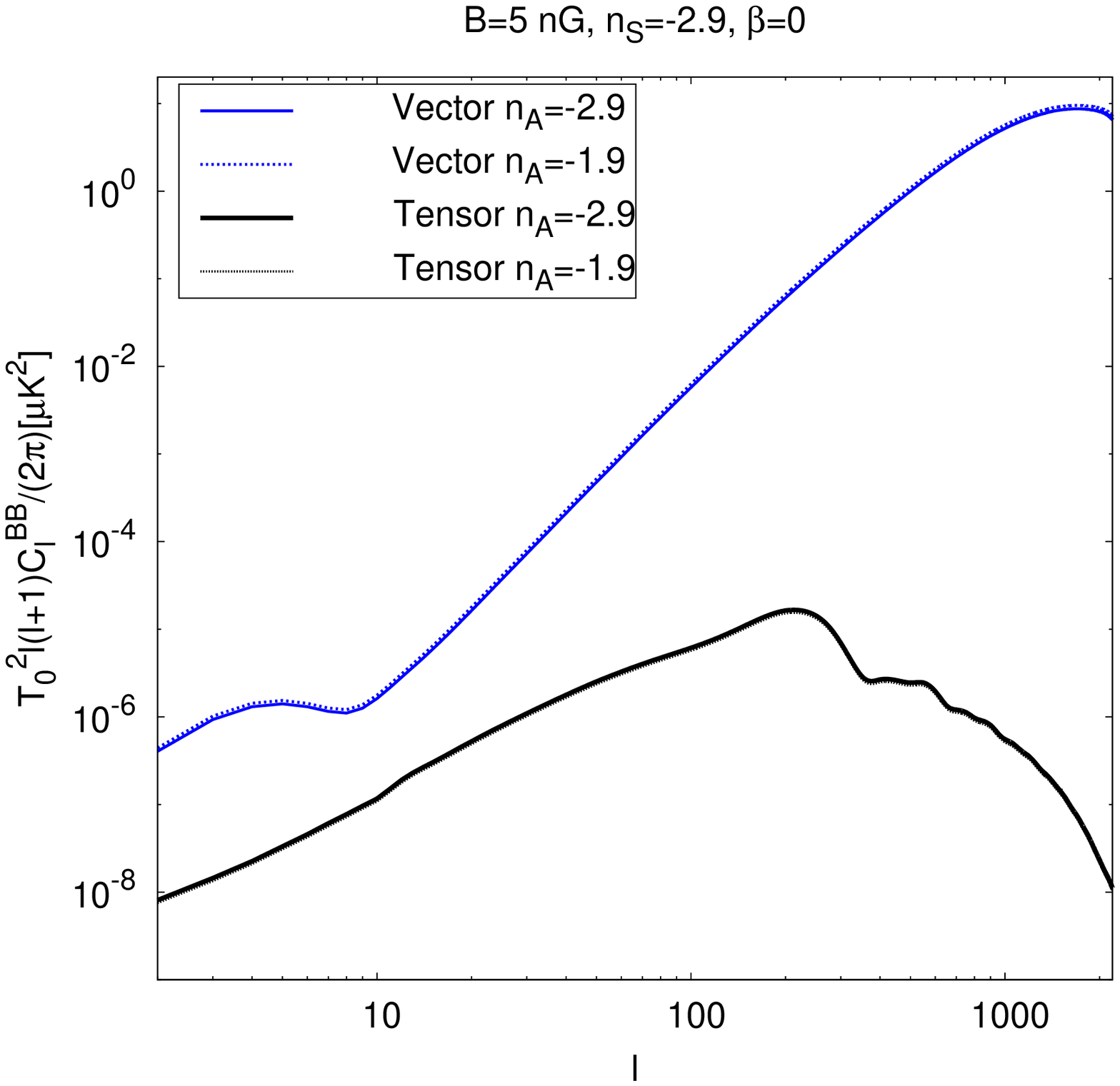}}
\caption{The TT,  EE,  TE and BB angular power spectra for the scalar, vector and tensor modes for the magnetic field strength $B=5$ nG, spectral index of the symmetric part of the magnetic field correlation function $n_S=-2.9$. The spectral index of the asymmetric part is assumed to be $n_A=-2.9$ and $n_A=-1.9$, respectively. The pure magnetic mode is shown, $\beta=0$.}
\label{fig4}
\end{figure}

\begin{figure}[h!]
\centerline{\epsfxsize=3.1in\epsfbox{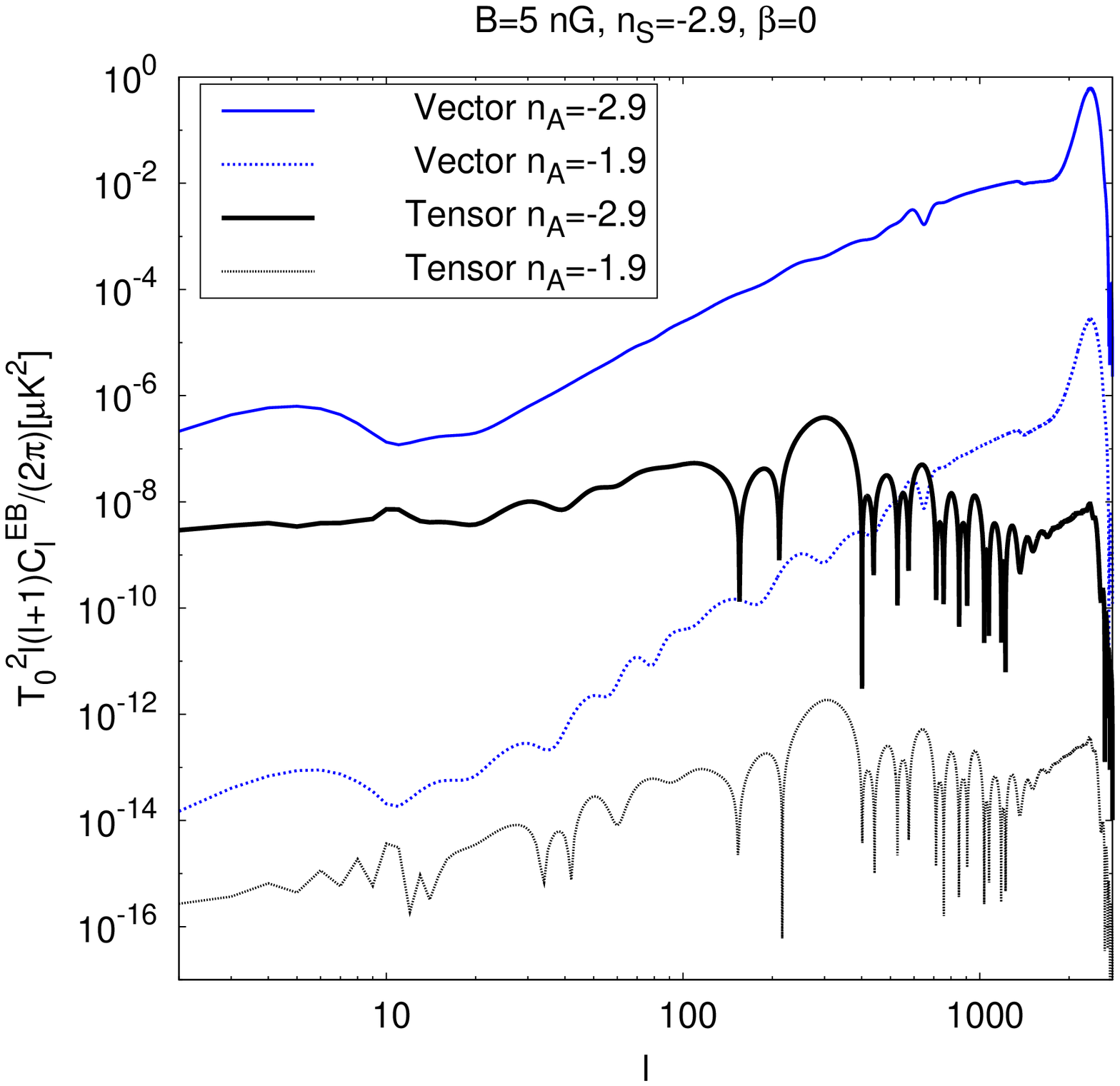}
\hspace{0.2cm}
\epsfxsize=3.1in\epsfbox{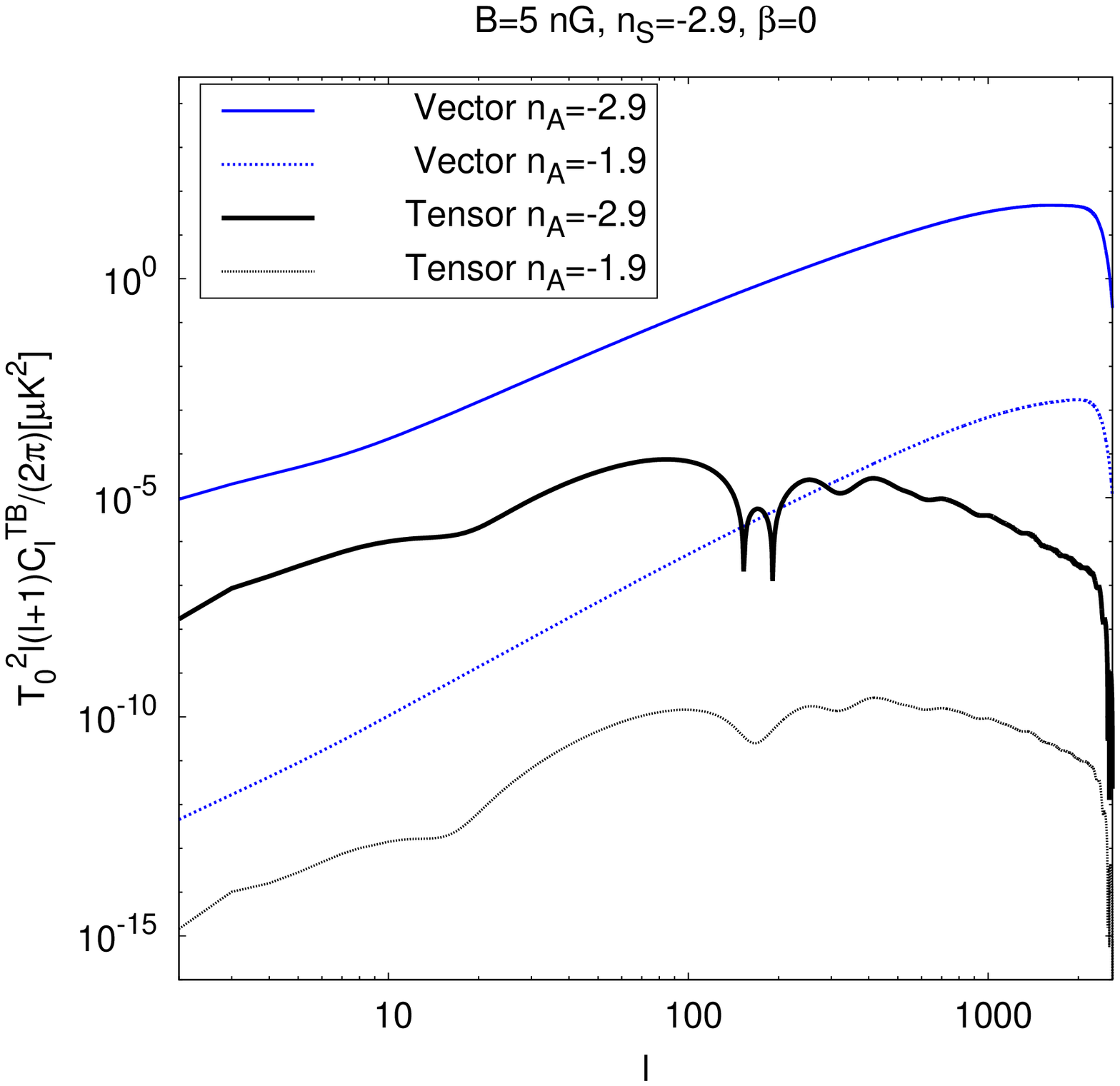}}
\caption{The angular power spectra $C_{\ell}^{EB}$ and $C_{\ell}^{TB}$  due to  vector and tensor modes for the magnetic field strength $B=5$ nG, spectral index of the symmetric part of the magnetic field correlation function $n_S=-2.9$. The spectral index of the asymmetric part is assumed to be $n_A=-2.9$ and $n_A=-1.9$, respectively. The pure magnetic mode is shown, $\beta=0$.}
\label{fig5}
\end{figure}

\begin{figure}[h!]
\centerline{\epsfxsize=3.1in\epsfbox{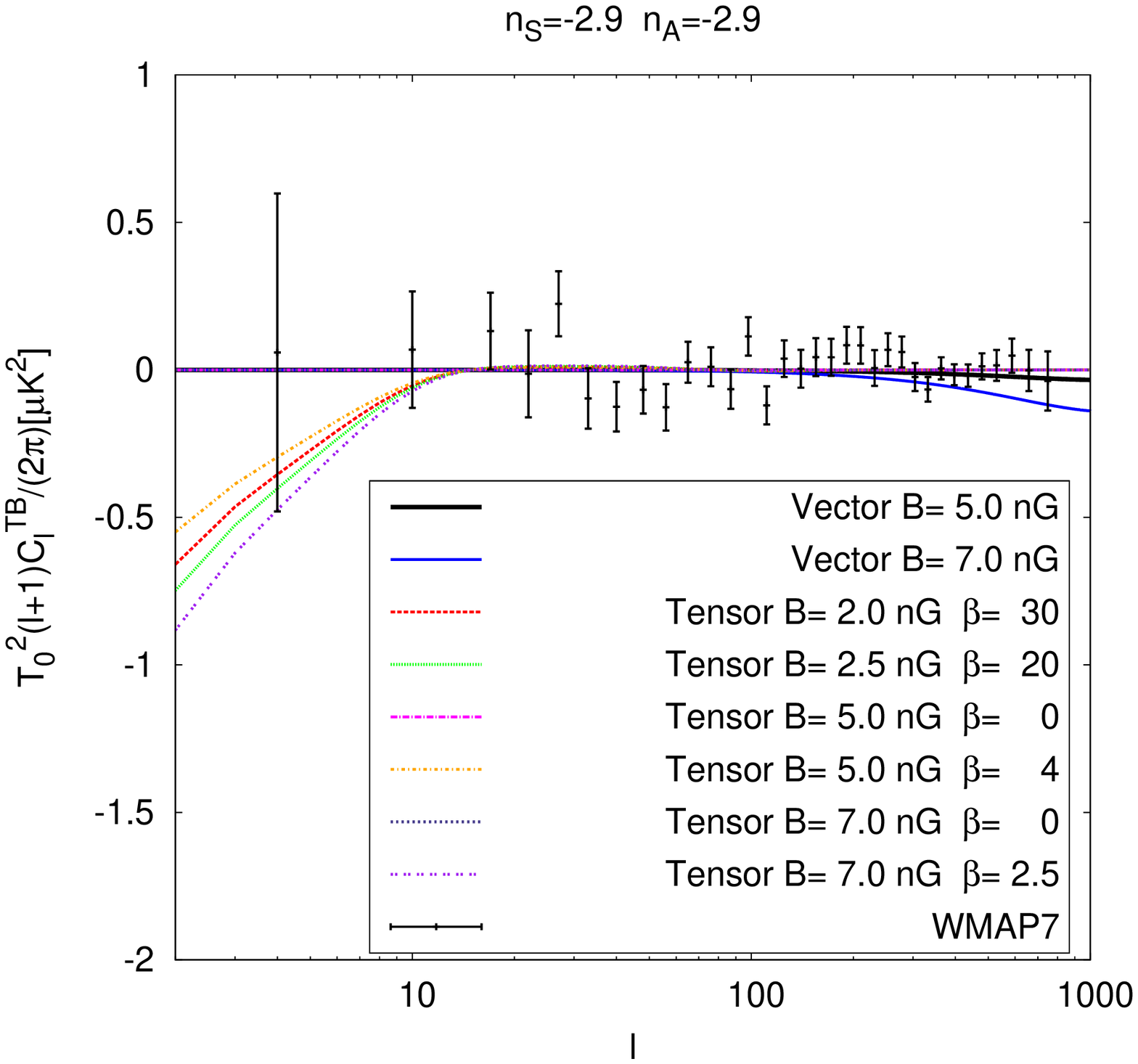}
\hspace{0.2cm}
\epsfxsize=3.1in\epsfbox{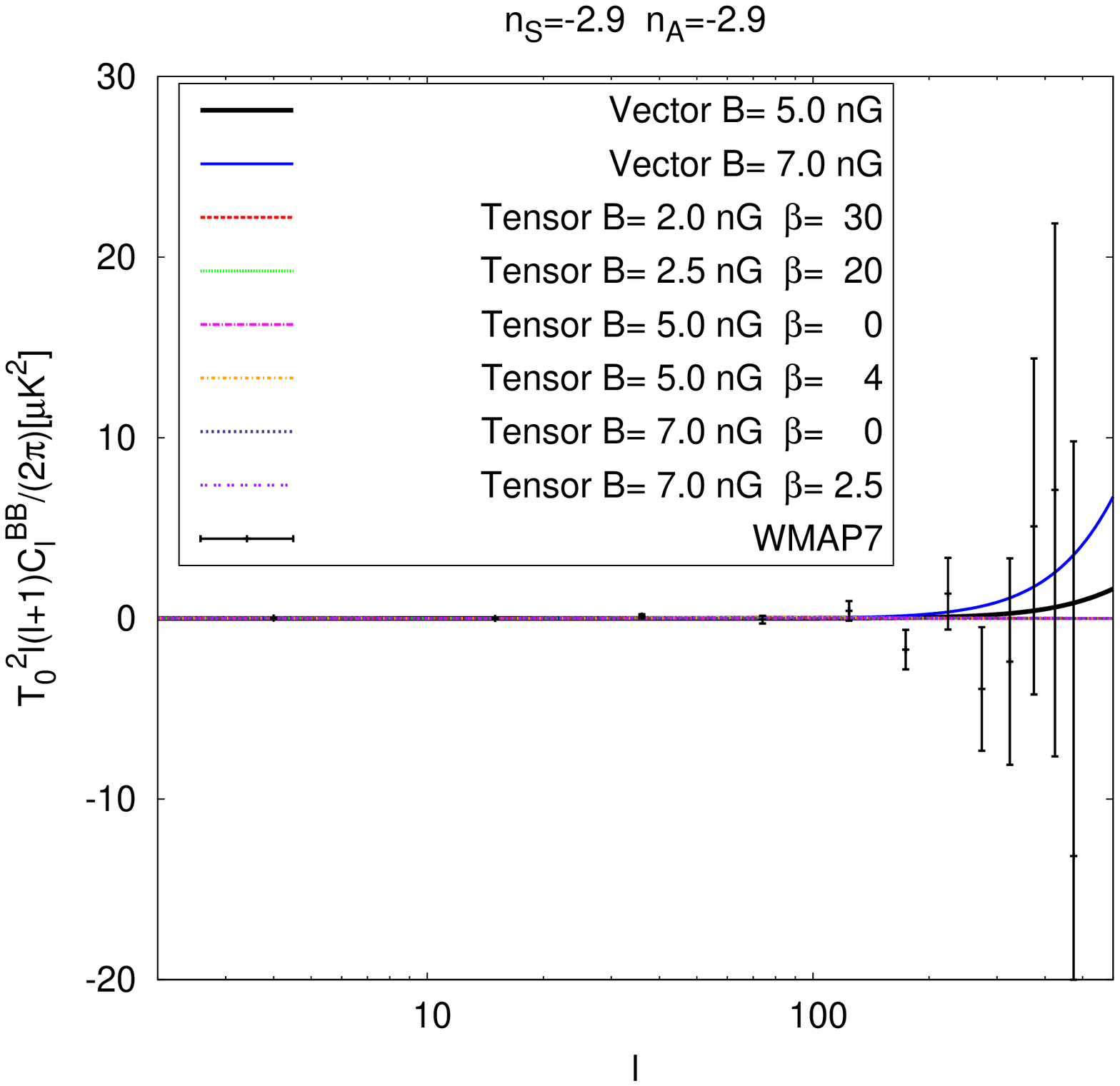}}
\caption{The TB and BB angular power spectra for different parameters comparing with WMAP7 data. Note that for TB the plotted angular power spectrum is 
$(\ell +1)C_{\ell}^{TB}/(2\pi)$ in accordance with the WMAP7 data and not $\ell(\ell+1)C_{\ell}^{TB}/(2\pi)$.}
\label{fig6}
\end{figure}

In figures \ref{fig4}-\ref{fig6} the angular power spectra determining the autocorrelation and cross correlation functions of the temperature anisotropies and polarization of the CMB  due to the  scalar, vector and tensor modes are shown for a choice of the magnetic field parameters and the best-fit values of the 6-parameter $\Lambda$CDM model of WMAP7 \cite{wmap7}, in particular, $\Omega_b=0.0445$, $\Omega_{\Lambda}=0.738$ and the reionization  optical depth 
$\tau=0.086$.  In all numerical solutions the amplitude of the helical part of the magnetic field is taken to be the maximally allowed value by the realizability condition (cf. equation (\ref{hb})). Moreover, for $n_A=-1.9$ the maximal wave number is set to $k_{max}=10^{2}k_m$.  The spectral indices of the symmetric and asymmetric parts are chosen to be negative. Whereas magnetic fields generated during inflation with cosmologically relevant field strengths result to have negative spectral indices (e.g. \cite{tw}) those generated during  a phase transition, such as e.g. the electroweak one, have to satisfy causality constraints which
require positive spectral indices \cite{cd,cdk}.  
The asymmetric part of magnetic field correlation function causes non vanishing cross correlations between the E-mode and B-mode and the temperature and the B-mode, respectively. These are shown in figure \ref{fig5}.
Current observations of the B-mode of polarization are consistent with zero \cite{wmap7,quad09}. Thus comparing in particular the cross correlation between temperature and the B-mode seems a promising possibility to  constrain the helical contribution. In figure \ref{fig6} the TB- and the BB- angular spectra have been plotted for different choices of parameters together with observational data from WMAP7 \cite{lambda}. 
As can be appreciated from figure \ref{fig6} the constraint on the contribution to $C_{\ell}^{TB}$ due to the vector mode is at high values of  the multipoles $\ell$
and for the tensor modes at low values of $\ell$. Whereas the vector modes do not depend on the parameter $\beta$ encoding the evolution of relevant quantities upto the time of neutrino decoupling, a nonzero value of $\beta$ constrains the magnetic field strength to lower values due to the tensor modes. 
Assuming that the magnetic field is created during inflation and thus setting $\tau_B$ to the beginning of the standard radiation dominated era, so that 
$\beta=\ln\frac{T_{RH}}{T_{\nu}}$, then assuming reheating at  $10^{10}$ GeV and together with $T_{\nu}=1$ MeV resulting in $\beta=30$, puts an upper limit on the  magnetic field strength of  $B=2$ nG for $n_S=n_A=-2.9$. For lower reheat temperatures larger values of the magnetic field strength are allowed.
The WMAP7 data for the BB spectrum are less constraining as can be seen in figure \ref{fig6} ({\it right}).
In the numerical solutions it was assumed that  there is no correlation between the magnetic field contributions and any primordial curvature perturbation or tensor modes from, e.g., inflation. However, if the magnetic field is generated during inflation then one might expect a correlation which deserves further study. This has recently been considered in \cite{cmk}.

\section{Conclusions}
\label{sect6}

The CMB anisotropies and polarization in the presence of a stochastic helical magnetic field have been calculated for scalar, vector and tensor modes.
For this purpose the modified version of CMBEASY  \cite{kek} calculating  the CMB anisotropies due to the scalar perturbations in the presence of a non helical 
magnetic field has been expanded. Firstly the numerical solution for the corresponding correlation functions has been included for scalar, vector and 
tensor modes. Secondly a new part has been added to include the calculation of the CMB anisotropies and polarization due to vector and tensor modes using the total angular momentum approach of Hu and White \cite{hw}.
A Gaussian window function is used to effectively cut-off the magnetic field spectrum at a wave number corresponding to the magnetic damping scale.
In the case of the scalar and tensor perturbations the initial conditions for the numerical solution which are set long after neutrino decoupling include 
a contribution encoding the evolution of relevant quantities due to the presence of the magnetic field before neutrino decoupling.
In the case of the scalar perturbations the comoving curvature perturbation grows upto the time of neutrino decoupling due the magnetic anisotropic stress. 
After neutrino decoupling the neutrino anisotropic stress compensates the magnetic anisotropic stress and the comoving curvature perturbation becomes a constant on superhorizon scales \cite{kkm,le,BC}. In the case of the tensor modes a similar behaviour is found for the amplitude of the tensor mode which has been shown here explicitly. The presence of a magnetic field prior to neutrino decoupling does not affect significantly the evolution of the vector modes.
Using a standard Boltzmann solver the contribution due to presence of the magnetic field before neutrino decoupling has been included in the initial conditions for  the numerical solution which are set long after neutrino decoupling.  However, since the presence of a magnetic field affects the evolution of the scalar and tensor perturbations before neutrino decoupling in order to be more precise one would have to start the numerical evolution and thus set the initial conditions for the Boltzmann code before neutrino decoupling. A problem we hope to address in the future.

In the case of a helical magnetic field in addition to the temperature (T) and polarization E- and B- mode autocorrelation spectra and cross correlation TE angular power spectrum there are also the cross correlation EB and TB angular power spectra.
The latter one has been used to illustrate the use of the current WMAP7 data to constrain the magnetic field parameters.

In comparison to earlier work on the effects of a helical magnetic field \cite{pvw,cdk,kr} on the CMB here a full numerical treatment has been provided including the correct initial conditions and evolution equations in the presence of a magnetic field as well as including numerical solutions for the different correlation functions of the magnetic field contributions. Moreover, the magnetic field spectrum  is effectively cut-off using a Gaussian window function.

\section{Acknowledgements}
I would like to thank the Kavli Institute for Cosmological Physics at the University of Chicago for hospitality where part of this work was done 
and Angela Olinto for interesting discussions. Also I am grateful to Juan Garcia-Bellido for drawing my attention to the problem of helical magnetic fields and the CMB.
Financial support by Spanish Science Ministry grants FPA2009-10612, FIS2009-07238 and CSD2007-00042 is gratefully acknowledged.
Furthermore I  acknowledge the use of the Legacy Archive for Microwave Background Data Analysis (LAMBDA). Support for LAMBDA is provided by the NASA Office of Space Science.


\begin{thebibliography}{99}

\bibitem{obsgal}
P.~P.~Kronberg,
  Rept.\ Prog.\ Phys.\  {\bf 57} (1994) 325;
 R. Wilebinski and R. Beck, Lect. Notes Phys. {\bf 664} (2005);
 C.~L.~Carilli and G.~B.~Taylor,
  Ann.\ Rev.\ Astron.\ Astrophys.\  {\bf 40} (2002) 319.


\bibitem{obscosmag}
I.~Vovk, A.~M.~Taylor, D.~Semikoz and A.~Neronov,
  ``Fermi/LAT observations of 1ES 0229+200: implications for extragalactic magnetic fields and background light,''
  arXiv:1112.2534 [astro-ph.CO];
A.~Neronov and I.~Vovk,
 Science {\bf 328} (2010) 73;
S.~'i.~Ando and A.~Kusenko,
  Astrophys.\ J.\  {\bf 722} (2010) L39.

  
\bibitem{reviews}
D.~Grasso, H.~R.~Rubinstein,
  Phys.\ Rept.\  {\bf 348 } (2001)  163;
L.~M.~Widrow,
  Rev.\ Mod.\ Phys.\  {\bf 74 } (2002)  775;
M.~Giovannini,
 Int.\ J.\ Mod.\ Phys.\ D {\bf 13} (2004) 391;
A.~Kandus, K.~E.~Kunze, C.~G.~Tsagas,
  Phys.\ Rept.\  {\bf 505 } (2011)  1;
L.~M.~Widrow, D.~Ryu, D.~Schleicher, K.~Subramanian, C.~G.~Tsagas and R.~A.~Treumann,
 ``The First Magnetic Fields,''
  arXiv:1109.4052 [astro-ph.CO];
 D.~Ryu, D.~R.~G.~Schleicher, R.~A.~Treumann, C.~G.~Tsagas and L.~M.~Widrow,
  ``Magnetic fields in the Large-Scale Structure of the Universe,''
  arXiv:1109.4055 [astro-ph.CO];
  
  
\bibitem{tw}
M.~S.~Turner and L.~M.~Widrow,
  Phys.\ Rev.\ D {\bf 37} (1988) 2743.


\bibitem{pt}
T.~Vachaspati,
  Phys.\ Lett.\ B {\bf 265} (1991) 258;
M.~Joyce and M.~E.~Shaposhnikov,
  Phys.\ Rev.\ Lett.\  {\bf 79} (1997) 1193;
J.~Ahonen and K.~Enqvist,
  Phys.\ Rev.\ D {\bf 57} (1998) 664;
  M.~M.~Forbes and A.~R.~Zhitnitsky,
  Phys.\ Rev.\ Lett.\  {\bf 85} (2000) 5268;
  A.~Diaz-Gil, J.~Garcia-Bellido, M.~Garcia Perez and A.~Gonzalez-Arroyo,
  Phys.\ Rev.\ Lett.\  {\bf 100} (2008) 241301;
 A.~Diaz-Gil, J.~Garcia-Bellido, M.~G.~Perez and A.~Gonzalez-Arroyo,
  JHEP {\bf 0807} (2008) 043.

 
 \bibitem{infhel}
  G.~B.~Field and S.~M.~Carroll,
  Phys.\ Rev.\ D {\bf 62} (2000) 103008;
  L.~Campanelli and M.~Giannotti,
  Phys.\ Rev.\ D {\bf 72} (2005) 123001;
    L.~Campanelli,
  Int.\ J.\ Mod.\ Phys.\ D {\bf 18} (2009) 1395;
R.~Durrer, L.~Hollenstein and R.~K.~Jain,
  JCAP {\bf 1103} (2011) 037.

\bibitem{pvw}
  L.~Pogosian, T.~Vachaspati, S.~Winitzki,
    Phys.\ Rev.\  {\bf D65 } (2002)  083502.


\bibitem{cdk}
  C.~Caprini, R.~Durrer, T.~Kahniashvili,
  Phys.\ Rev.\  {\bf D69 } (2004)  063006.


\bibitem{kr}
T.~Kahniashvili and B.~Ratra,
  Phys.\ Rev.\ D {\bf 71} (2005) 103006.


\bibitem{cosmics}
E.~Bertschinger,
  ``COSMICS: cosmological initial conditions and microwave anisotropy codes,''
  astro-ph/9506070.


\bibitem{cmbfast}
U.~Seljak and M.~Zaldarriaga,
  Astrophys.\ J.\  {\bf 469} (1996) 437;
  \newline
http://lambda.gsfc.nasa.gov/toolbox/tb\underline{ }cmbfast\underline{ }ov.cfm.


\bibitem{camb}
A.~Lewis, A.~Challinor and A.~Lasenby,
 Astrophys.\ J.\  {\bf 538} (2000) 473;
 \newline 
http://camb.info/.


\bibitem{cmbeasy}
M.~Doran,
JCAP {\bf 0510} (2005) 011;
  \newline
http://www.thphys.uni-heidelberg.de/$\tilde{\;}$robbers/cmbeasy/


\bibitem{class}
J.~Lesgourgues,
 ``The Cosmic Linear Anisotropy Solving System (CLASS) I: Overview,''
  arXiv:1104.2932 [astro-ph.IM].
\newline  
http://lesgourg.web.cern.ch/lesgourg/class.php


\bibitem{grant}
D.~Yamazaki, K.~Ichiki, T.~Kajino and G.~J.~Mathews,
   Astrophys.\ J.\  {\bf 646} (2006) 719;
   D.~Yamazaki, K.~Ichiki, T.~Kajino and G.~J.~Mathews,
  Phys.\ Rev.\  D {\bf 77} (2008) 043005;
  K.~Kojima, K.~Ichiki, D.~G.~Yamazaki, T.~Kajino and G.~J.~Mathews,
  Phys.\ Rev.\ D {\bf 78} (2008) 045010.


\bibitem{GK}
 M.~Giovannini and K.~E.~Kunze,
  Phys.\ Rev.\  D {\bf 77} (2008) 061301;
   M.~Giovannini and K.~E.~Kunze,
Phys.\ Rev.\  D {\bf 77} (2008) 063003.


\bibitem{fin}
  F.~Finelli, F.~Paci and D.~Paoletti,
  Phys.\ Rev.\  D {\bf 78} (2008) 023510.
 D.~Paoletti, F.~Finelli and F.~Paci,
  Mon.\ Not.\ Roy.\ Astron.\ Soc.\  {\bf 396} (2009) 523;
  D.~Paoletti and F.~Finelli,
  Phys.\ Rev.\ D {\bf 83} (2011) 123533.


\bibitem{le}
  J.~R.~Shaw and A.~Lewis,
  Phys.\ Rev.\  D {\bf 81} (2010) 043517.
  

\bibitem{kek}
 K.~E.~Kunze,
  Phys.\ Rev.\  {\bf D83 } (2011)  023006.


\bibitem{hw}
W.~Hu, M.~J.~White,
 Phys.\ Rev.\  {\bf D56 } (1997)  596.


\bibitem{ks}
 H.~Kodama and M.~Sasaki,
  Prog.\ Theor.\ Phys.\ Suppl.\  {\bf 78} (1984) 1.


\bibitem{wmap7}
D.~Larson, J.~Dunkley, G.~Hinshaw, E.~Komatsu, M.~R.~Nolta, C.~L.~Bennett, B.~Gold and M.~Halpern {\it et al.},
 Astrophys.\ J.\ Suppl.\  {\bf 192} (2011) 16.
  
  
  \bibitem{sb}
 K.~Subramanian and J.~D.~Barrow,
  Phys.\ Rev.\  D {\bf 58} (1998) 083502.


\bibitem{tsagas}
C.~G.~Tsagas and J.~D.~Barrow,
  Class.\ Quant.\ Grav.\  {\bf 14} (1997) 2539;
  C.~G.~Tsagas and J.~D.~Barrow,
 Class.\ Quant.\ Grav.\  {\bf 15} (1998) 3523;
 J.~D.~Barrow, R.~Maartens and C.~G.~Tsagas,
  Phys.\ Rept.\  {\bf 449} (2007) 131.

  
\bibitem{Thorne}
K.~S.~Thorne,
Rev.\ Mod.\ Phys.\  {\bf 52} (1980) 299.


\bibitem{Rose}
M. E. Rose, {\it Elementary Theory of Angular Momentum} (Dover Publications, INC., New York, USA, 1957).


\bibitem{bis1}
D. Biskamp, {\it Nonlinear Magnetohydrodynamics}, (Cambridge University Press, Cambridge, UK, 1997).


\bibitem{bis2}
D. Biskamp, {\it Magnetohydrodynamic Turbulence}, (Cambridge University Press, Cambridge, UK, 2003).


\bibitem{fa}
J. H. Finn and T. M. Antonsen, 
Comments Plasma Phys. Controlled Fusion {\bf 9} (1985) 111. 


\bibitem{sbra}
 K.~Subramanian, A.~Brandenburg,
  Astrophys.\ J.\  {\bf 648 } (2006)  L71.


\bibitem{mb}
 L.~Malyshkin, S.~Boldyrev,
  Astrophys.\ J.\  {\bf 671 } (2007)  L185.


\bibitem{my2}
A.S. Monin and A.M. Yaglom, {\it Statistical Fluid Mechanics: Mechanics of Turbulence, Volume II}, (Dover Publications Inc., NewYork, USA, 1975).


\bibitem{sub1}
K.~Subramanian,
  Phys.\ Rev.\ Lett.\  {\bf 83 } (1999)  2957.


\bibitem{sigl}
 G.~Sigl,
  Phys.\ Rev.\  {\bf D66 } (2002)  123002.


\bibitem{ab}
   A.~Brandenburg,
  ``The critical role of magnetic helicity in astrophysical large-scale dynamos,''
  [arXiv:0909.4377 [astro-ph.SR]].


\bibitem{kko}
K.~Jedamzik, V.~Katalinic and A.~V.~Olinto,
  Phys.\ Rev.\ D {\bf 57} (1998) 3264.

  
 \bibitem{kkm}
 K.~Kojima, T.~Kajino and G.~J.~Mathews,
  JCAP {\bf 1002} (2010) 018.


\bibitem{BC}
 C.~Bonvin and C.~Caprini,
  JCAP {\bf 1005} (2010) 022.
   
   
  \bibitem{mg}
M.~Giovannini,
  Phys.\ Rev.\  {\bf D74 } (2006)  063002.
 
   
 \bibitem{py}
P.~J.~E.~Peebles, J.~T.~Yu,
  Astrophys.\ J.\  {\bf 162 } (1970)  815.


\bibitem{maber}
C.~-P.~Ma, E.~Bertschinger,
  Astrophys.\ J.\  {\bf 455 } (1995)  7.


\bibitem{doran}
M.~Doran,
  JCAP {\bf 0506 } (2005)  011.


\bibitem{bcd}
C.~Bonvin, C.~Caprini and R.~Durrer,
  ``Magnetic fields from inflation: the fatal transition to the radiation era,''
  arXiv:1112.3901 [astro-ph.CO];
C.~Bonvin, C.~Caprini and R.~Durrer,
  ``Inflationary magnetic fields spoil the homogeneity and isotropy of the Universe,''
  arXiv:1112.3897 [astro-ph.CO].


\bibitem{iso}
 J.~Valiviita and V.~Muhonen,
   Phys.\ Rev.\ Lett.\  {\bf 91} (2003) 131302;
H.~Kurki-Suonio, V.~Muhonen and J.~Valiviita,
  Phys.\ Rev.\  D {\bf 71} (2005) 063005.


\bibitem{cd}
R.~Durrer and C.~Caprini,
 JCAP {\bf 0311} (2003) 010.

\bibitem{quad09}
  M.~L.~Brown {\it et al.}  [QUaD Collaboration],
  Astrophys.\ J.\  {\bf 705} (2009) 978.
  
 \bibitem{lambda}
  Legacy Archive for Microwave Background Data Analysis (LAMBDA), 
  http://lambda.gsfc.nasa.gov/
  
  
  \bibitem{cmk}
   R.~R.~Caldwell, L.~Motta and M.~Kamionkowski,
  ``Correlation of inflation-produced magnetic fields with scalar fluctuations,''
  arXiv:1109.4415 [astro-ph.CO].


\end{thebibliography}
\end{document}